\documentclass[fleqn,usenatbib]{mnras}

\usepackage{newtxtext,newtxmath}

\usepackage{pdflscape}	
\usepackage{longtable} 

\usepackage[T1]{fontenc}
\usepackage{xcolor}
\usepackage{tcolorbox}
\usepackage{orcidlink}

\DeclareRobustCommand{\VAN}[3]{#2}
\let\VANthebibliography\thebibliography
\def\thebibliography{\DeclareRobustCommand{\VAN}[3]{##3}\VANthebibliography}

\newcommand{\ACTrevision}[1]{{\textcolor{black}{#1}}}

\usepackage{graphicx}	
\usepackage{amsmath}	
\usepackage{subfigure}

\title[Optical Cluster Miscentering from SZ Signals]{Miscentering of Optical Galaxy Clusters Based on Sunyaev-Zeldovich Counterparts}

\author[Ding et al.]{
Jupiter Ding,$^{1, 2}$\thanks{E-mail: jupiterding2029@u.northwestern.edu}
Roohi Dalal,$^{1, 3, 4}$
Tomomi Sunayama,$^{5}$ 
Michael A. Strauss,$^{1}$
Masamune Oguri,$^{6, 7}$
\newauthor
Nobuhiro Okabe,$^{8}$
Matt Hilton,$^{9, 10}$
Rogério Monteiro-Oliveira,$^{11}$
Crist\'obal Sif\'on,$^{12}$
Suzanne T. Staggs$^{13}$
\\
$^{1}$Department of Astrophysical Sciences, Princeton University, Peyton Hall, Princeton, NJ 08544, USA\\
$^{2}$CIERA and Department of Physics and Astronomy, Northwestern University, 2145 Sheridan Road, Evanston, IL 60208, USA\\
$^{3}$Department of Physics and Astronomy, University of British Columbia, Vancouver, British Columbia, Canada.\\
$^{4}$Department of Political Science, University of British Columbia, Vancouver, British Columbia, Canada\\
$^{5}$Department of Astronomy and Steward Observatory, University of Arizona, Tucson, AZ 85721, USA\\
$^{6}$Center for Frontier Science, Chiba University, 1-33 Yayoi-cho, Inage-ku, Chiba 263-8522, Japan\\
$^{7}$Department of Physics, Graduate School of Science, Chiba University, 1-33 Yayoi-Cho, Inage-Ku, Chiba 263-8522, Japan\\
$^{8}$Physics Program, Graduate School of Advanced Science and Engineering, Higashi Hiroshima Campus, Hiroshima University, Hiroshima 739-8526, Japan\\
$^{9}$ Wits Centre for Astrophysics, School of Physics, University of the Witwatersrand, Private Bag 3, 2050, Johannesburg, South Africa\\
$^{10}$ Astrophysics Research Centre, School of Mathematics, Statistics, and Computer Science, University of KwaZulu-Natal, Westville Campus,\\ Durban 4041, South Africa\\
$^{11}$ Institute of Astronomy and Astrophysics, Academia Sinica, Taipei 10617, Taiwan\\
$^{12}$ Instituto de F\'isica, Pontificia Universidad Cat\'olica de Valpara\'iso, Casilla 4059, Valpara\'iso, Chile\\
$^{13}$ Joseph Henry Laboratories of Physics, Jadwin Hall, Princeton University, Princeton, NJ, USA 08544
}

\date{Accepted XXX. Received YYY; in original form ZZZ}

\pubyear{2024}

\begin{document}
\label{firstpage}
\pagerange{\pageref{firstpage}--\pageref{lastpage}}
\maketitle

\begin{abstract}
The ``miscentering effect,'' i.e., the offset between a galaxy cluster's optically-defined center and the center of its gravitational potential, is a significant systematic effect on brightest cluster galaxy (BCG) studies and cluster lensing analyses. We perform a cross-match between the optical cluster catalog from the Hyper Suprime-Cam (HSC) Survey S19A Data Release and the Sunyaev-Zeldovich cluster catalog from Data Release 5 of the Atacama Cosmology Telescope (ACT). We obtain a sample of 186 clusters in common in the redshift range $0.1 \leq z \leq 1.4$ over an area of 469 deg$^2$. By modeling the distribution of centering offsets in this fiducial sample, we find a miscentered fraction (corresponding to clusters offset by more than \ACTrevision{330} kpc) of $\sim 25\%$, a value consistent with previous miscentering studies. We examine the image of each miscentered cluster in our sample and identify one of several reasons to explain the miscentering. Some clusters show significant miscentering for astrophysical reasons, i.e., ongoing cluster mergers.  Others are miscentered due to non-astrophysical, systematic effects in the HSC data or the cluster-finding algorithm. After removing all clusters with clear, non-astrophysical causes of miscentering from the sample, we find a considerably smaller miscentered fraction, $\sim 10\%$. We show that the gravitational lensing signal within 1 Mpc of miscentered clusters is considerably smaller than that of well-centered clusters, and we suggest that the ACT SZ centers are a better estimate of the true cluster potential centroid.   
\end{abstract}




\section{Introduction}\label{sec:intro}

As the most massive gravitationally self-bound objects in the universe, galaxy clusters are remarkably useful probes of both galaxy evolution and cosmology. The formation and evolution of brightest cluster galaxies (BCGs), the most luminous and massive galaxies in the universe, are likely influenced by merger processes and other astrophysical phenomena within their host clusters \citep{hoessel_ccd_1985, postman_brightest_1995, lauer_brightest_2014, huang_individual_2018, dalal_brightest_2021}. Moreover, clusters are strong probes of the growth of structure in the universe, and they can be used to constrain cosmological parameters through abundance studies \citep{bahcall_most_1998, allen_cosmological_2011, weinberg_observational_2013, mantz_weighing_2015, garrel_xxl_2022, clerc_x-ray_2023}. Measuring the galaxy weak lensing signal around a cluster \cite[e.g.,][]{rozo_cosmological_2010, abbott_dark_2020, sunayama_optical_2023} adds information about the halo mass to the abundance measurements, allowing one to simultaneously place strong constraints on both cluster masses and cosmology. 

Methods for observing or identifying clusters are varied; here, we describe three. Each of these methods relies on a different observational probe to determine the bottom of the cluster's gravitational potential well; in other words, its gravitational ``center.'' The first method we describe relies on the Sunyaev-Zeldovich (SZ) effect \citep{sunyaev_observations_1972}: electrons in a cluster's hot intracluster medium (ICM) can scatter photons from the Cosmic Microwave Background (CMB) to higher energies via inverse Compton scattering. As a result, in CMB maps taken at frequencies below 220 GHz, clusters appear as ``cold'' or darkened spots. Such a signal traces the cluster's gas---more precisely, the SZ signal at a specific position on the sky is proportional to the integrated gas pressure along the line of sight within the cluster. The cluster's SZ center is then defined as the centroid of the SZ signal. Since the cluster's gas should generally trace the mass distribution, the SZ center should coincide with the bottom of the cluster's potential well, especially when the cluster is fully virialized \citep{kravtsov_formation_2012}. 

A second method for identifying clusters relies on optical data. Optical surveys identify clusters by looking for closely concentrated collections of galaxies on the sky located at similar redshifts \cite[e.g.,][]{kim_detecting_2002, gladders_red-sequence_2005, yang_galaxy_2007, oguri_cluster_2014, rykoff_redmapper_2014}, as measured either photometrically or with a spectroscopic survey. Some cluster-finding algorithms define a cluster's optical center as the location of the ``central galaxy;'' i.e., a bright, massive galaxy located near the peak of the cluster's galaxy distribution. This is because the central galaxy often coincides with the bottom of its cluster's gravitational potential well as a result of dynamical friction \cite[][]{chandrasekhar_dynamical_1943}: when a relatively massive galaxy moves within a cluster, it experiences a gravitational drag force from the surrounding matter, causing it to spiral in towards the gravitational center \citep{ostriker_another_1975, hausman_galactic_1978, binney_galactic_2008}.

However, there are situations where the central galaxy does not coincide with the gravitational center. For example, consider two clusters in the process of merging. If they are close enough to each other, an optical cluster-finding algorithm may identify the system as one cluster even if it has not yet reached gravitational equilibrium; a typical relaxation time for a cluster merger is a few billion years \citep{faltenbacher_oscillatory_2006, machado_simulating_2015}. Consequently, there may not yet be a bright, massive galaxy sufficiently close to the gas profile center of the merged cluster. Alternatively, if the merging cluster is not yet fully virialized, there would not even be a well-defined bottom of the potential well. In both cases, one would be unable to accurately identify the cluster center using a central galaxy.

Lastly, X-ray cluster surveys \cite[e.g.,][]{ebeling_rosat_1998, pacaud_xmm-lss_2007, liu_erosita_2022}, similar to SZ cluster surveys, can trace the hot gas in a cluster: the ICM of a cluster emits X-rays via bremsstrahlung (or free-free) radiation.

Previous miscentering studies have generally shown that, around 20-40\% of the time, the central galaxy is \textit{not} a good indicator of the bottom of a cluster's potential well \citep{sehgal_atacama_2013, lauer_brightest_2014, rykoff_redmapper_2016, oguri_optically-selected_2018, hollowood_chandra_2019, zhang_dark_2019, bleem_sptpol_2020, yan_analysis_2020}. This number is referred to as the \textit{miscentered fraction}.\footnote{We will also use the term \textit{well-centered fraction}, which refers to the fraction of clusters for which the central galaxy \textit{is} a good indicator of the gravitational center.} In such studies, including this one, it is assumed that the cluster center derived from a gas profile (i.e., from SZ or X-ray data) is a good indicator of the bottom of the potential well, while a cluster's optical center may be displaced from the bottom of the potential well.

Miscentering is a major systematic effect in cluster lensing analyses which constrain cosmological parameters with the cluster mass function \citep{mcclintock_dark_2019}. Such studies take the lensing signals of optical clusters, model their azimuthally-averaged mass profiles, and combine this with an inference of cluster abundance as a function of some optically-based variable, e.g., richness. Richness is a measure of the number of galaxies in a cluster above some luminosity or mass threshold; we denote it in this paper as $\lambda$. (Note that the CAMIRA catalog, discussed in Section~\ref{subsubsec:hsc_cat}, denotes richness with $N_{\mathrm{mem}}$.) Miscentering of clusters impacts this measurement in several ways: first, the azimuthally averaged cluster lensing signal is suppressed because one is not centering the lensing signal on the true center of the cluster's mass profile. Furthermore, miscentering leads to the underestimation of the number of clusters above a certain richness threshold \citep{zhang_dark_2019}. Therefore, cluster lensing analyses explicitly account for the miscentered fraction in order to obtain estimates of cosmological parameters \cite[e.g.,][]{rozo_cosmological_2010, abbott_dark_2020, sunayama_optical_2023, kelly_dark_2023}. Additionally, BCG studies often define the BCG as the cluster's central galaxy \cite[e.g.,][]{oguri_optically-selected_2018}, so miscentering can introduce a bias in BCG samples. For example, \cite{dalal_brightest_2021} showed that BCGs with large displacements from the cluster center tend to be less dominant than BCGs that are not. Finally, miscentering studies can point us toward interesting astrophysics within clusters (e.g., mergers) that contribute to miscentering (and may also influence BCG evolution). 

In this paper, we aim to evaluate potential causes of the miscentering effect. As discussed previously, there may be astrophysical processes like mergers that cause central galaxies to be offset from the gas profile center. We also explore non-astrophysical causes of miscentering, which include systematic effects in optical cluster data. Broadly speaking, our work illuminates potential directions for improvement in optical cluster-finding algorithms. In the context of BCG studies and cluster cosmology, we hope that identifying causes of miscentering will allow for more effective quantification and mitigation of this systematic effect.

This paper is structured as follows: in Section~\ref{sec:data}, we describe our cross-match between an optical cluster catalog, drawn from the imaging data of the Hyper Suprime-Cam (HSC) survey, and an SZ cluster catalog from the Atacama Cosmology Telescope (ACT) survey.  The result is a matched sample that is larger and deeper than those used in many previous miscentering studies. In Section~\ref{sec:offset_model}, we model the distribution of centering offsets within the fiducial sample and infer the fraction of clusters that are miscentered. In Section~\ref{sec:causes_of_miscent}, we examine the optical image of each miscentered cluster and identify one of several potential causes of the miscentering. We then remove clusters with non-astrophysical causes of miscentering from our analysis and characterize the distribution of offsets among the remaining clusters. We summarize the results, potential implications, and avenues for future work in Section~\ref{sec:discussion}. Throughout our analysis, we assume a $\Lambda$CDM cosmology with $H_0=70$ km/s/Mpc or $h=0.7$, $\Omega_{m,0} = 0.3$, and $\Omega_{\Lambda, 0}=0.7$.

\section{Data}\label{sec:data}

\begin{figure}
     \centering
     \includegraphics[width=0.48\textwidth]{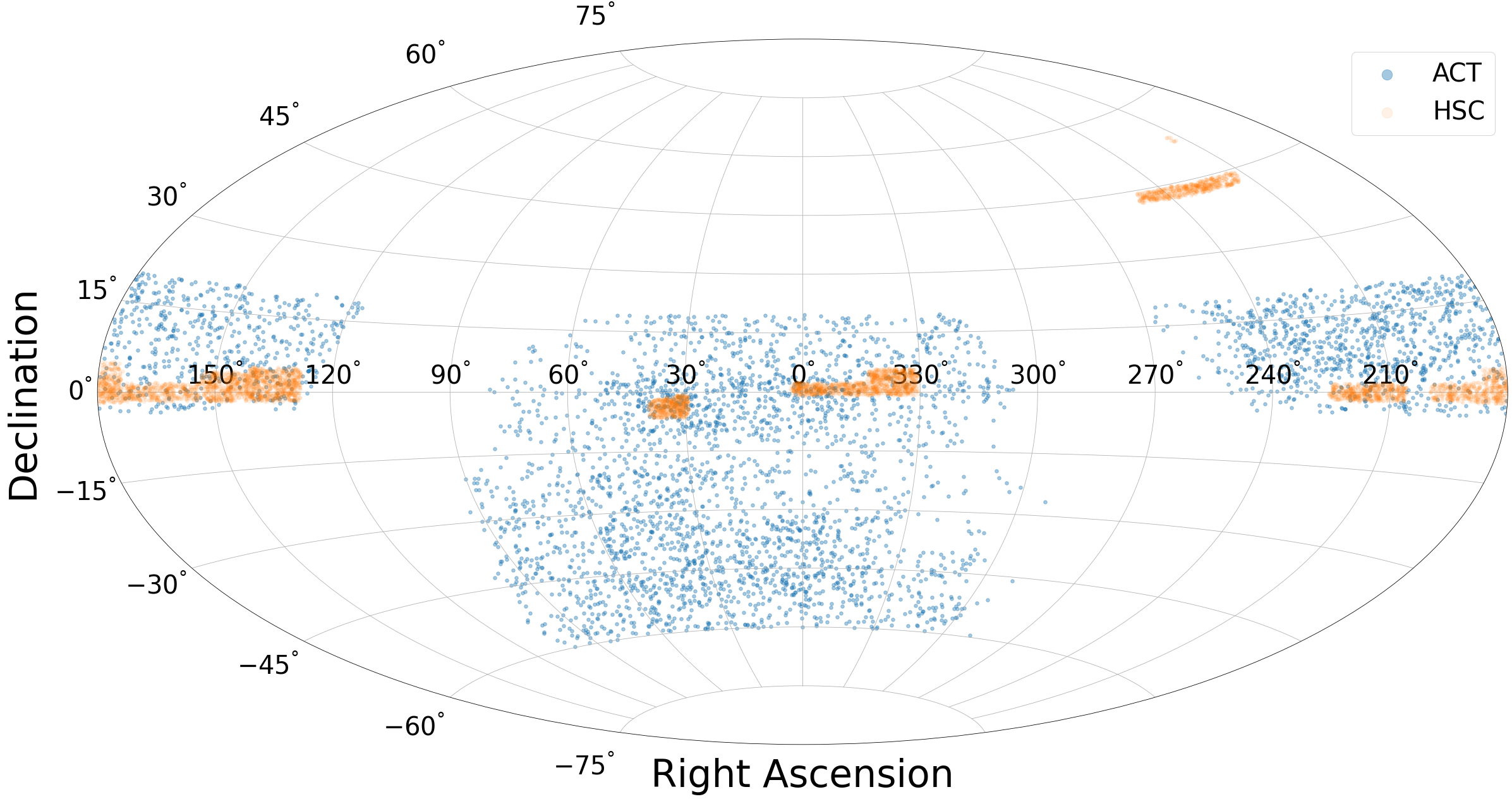}
     \caption{Sky map, in equatorial coordinates, of the positions of the clusters in the HSC and ACT catalogs used to create the fiducial sample. In total, we considered 5860 HSC clusters with $\lambda > 15$ and 4195 ACT clusters. The overlap in area between these two catalogs is 469 $\text{deg}^2$.}
     \label{fig:cls_on_sky}
\end{figure}

\subsection{Cluster Catalogs Used}\label{subsec:catalogs}

In this study, we cross-match an optical cluster catalog \citep{oguri_optically-selected_2018} from the Hyper Suprime-Cam (HSC) Subaru Strategic Program \citep{aihara_hyper_2018} with an SZ cluster catalog from the Atacama Cosmology Telescope (ACT) \citep{hilton_atacama_2021}. In particular, we use the optical catalog from the HSC S19A data release \citep{aihara_third_2022} along with the SZ catalog from Data Release 5 of ACT (DR5; \citealt{naess_atacama_2020}); both catalogs are described in detail below. Fig.~\ref{fig:cls_on_sky} shows the distribution on the sky of the clusters considered in the cross-match.

\subsubsection{HSC Catalog}\label{subsubsec:hsc_cat}

The Hyper Suprime-Cam is a wide-field optical imaging camera installed on the Subaru 8.2-m telescope \citep{miyazaki_hyper_2018}. The HSC survey \citep{aihara_hyper_2018} has three layers---Wide, Deep, and Ultradeep. In the completed survey, the HSC Wide layer, which we use in this work, has observed a sky area of $\sim 1200 \text{ deg}^2$ with five broadband filters (\textit{grizy}; \citealt{kawanomoto_hyper_2018}). The HSC survey is unprecedented in that its depth ($i_\text{lim} \sim 26$ at $5\sigma$ for point sources) is sufficient to construct a large sample of optically-observed clusters up to $z \sim 1.4$ \citep{oguri_optically-selected_2018}. In this work, we use the HSC S19A wide-field, star-masked catalog (the star masks used in the HSC survey are discussed in Section~\ref{subsec:hsc_effects}.) The S19A data release of the HSC survey consists of data taken between March 2014 and April 2019, covering an area of $\sim 589 \text{ deg}^2$. 

The cluster catalog itself is created via the CAMIRA (Cluster-finding Algorithm based on Multi-band Identification of Red-sequence gAlaxies) algorithm \citep{oguri_cluster_2014, oguri_optically-selected_2018}. In short, CAMIRA searches for overdensities of red-sequence galaxies that are similar in both photometric redshift and position on the sky. CAMIRA identifies the cluster's center as the position of a massive galaxy near the peak of the cluster's galaxy distribution; this is often, but not always, the \textit{brightest cluster galaxy} (BCG), or the most luminous galaxy in the cluster. Additionally, CAMIRA uses stellar population synthesis models to compute a cluster's photometric redshift from the cluster's high-confidence member galaxies. The S19A CAMIRA catalog consists of clusters in the redshift range $0.1 \leq z \leq 1.38$.

We emphasize that BCGs and central galaxies (as described in Section~\ref{sec:intro}) are \textit{not} always interchangeable terms. The definition of BCG varies between studies: for example, \cite{lauer_brightest_2014} and \cite{dalal_brightest_2021} use the literal definition of BCG (namely, the brightest galaxy in the cluster), while \cite{oguri_cluster_2014} defines a BCG as the most luminous galaxy among galaxies close to the peak of the cluster's galaxy distribution. Hence, \cite{oguri_cluster_2014} uses BCG and central galaxy interchangeably. For the sake of clarity, we follow \cite{dalal_brightest_2021} and avoid using the term ``BCG'' when referring to the central galaxy.

One should also note that the task of identifying an optical cluster---including its member galaxies and center---is not straightforward. In trying to identify a candidate central galaxy as a cluster's center, one must have a general understanding of which galaxies belong to the cluster. However, in deciding which galaxies belong to a cluster, one must consider how close these galaxies are to the cluster center. Therefore, identifying the optical center is an inherently \ACTrevision{iterative} process. In practice, CAMIRA uses a likelihood function to determine the most likely central galaxy for a given cluster candidate and then assigns a membership probability to each galaxy assigned to the cluster. This probability is based in part on its distance from the central galaxy, where the central galaxy has a membership probability of 1.

\subsubsection{ACT Catalog}\label{subsubsec:act_cat}

The SZ catalog presented here comes from ACT Data Release 5; the catalog is constructed from ACT observations obtained from 2008 to 2018 \citep{naess_atacama_2020}. The catalog covers 13,211 $\text{deg}^2$ of the sky, with an overlap of 469 $\text{deg}^2$ with the HSC S19A wide survey. To search for clusters, \ACTrevision{\cite{hilton_atacama_2021} apply a multi-frequency matched filter simultaneously to two sky maps from \cite{naess_atacama_2020} --- one each at 90 and 150 GHz}. The filter has a spatial component, which searches for cluster-sized objects, and a spectral component, which weights the signal according to the known spectral dependency of the SZ effect. Afterwards, \cite{naess_atacama_2020} compile a list of the SNR peaks; these are the cluster candidates \citep{hilton_nemo_2023}. To create the final cluster catalog, \cite{hilton_atacama_2021} match their candidates with optical imaging surveys, including the Sloan Digital Sky Survey (SDSS) \citep{ahumada_16th_2020}, the Dark Energy Survey (DES) \citep{abbott_dark_2018}, and HSC, in addition to estimating photometric redshifts by applying the \texttt{zCluster} algorithm \citep{hilton_zcluster_2023} to photometric data from the Dark Energy Camera Legacy Survey (DECaLS, \citealt{dey_overview_2019}), the Kilo Degree Survey (KiDS, \citealt{wright_kidsviking-450_2019}), and SDSS. 

In this work, we perform an independent cross-match between the final list of ACT candidates and the HSC catalog (see Section~\ref{subsec:cross_match_sample}). This is because the ACT cluster dataset \citep{hilton_atacama_2021} does not include HSC cluster data for each ACT cluster with an HSC match; instead, some of these clusters are listed with optical data based on other surveys. We did not, however, cross-check the initial list of ACT cluster candidates with the HSC catalog. For $\text{SNR}_{2.4} \gtrsim 5.0$,\footnote{As defined in \cite{hilton_atacama_2021}, $\text{SNR}_{2.4}$ is the SZ signal's signal-to-noise ratio at \cite{hilton_atacama_2021}'s reference 2.4' filter scale.} more than 95\% of ACT cluster candidates within the HSC footprint are optically confirmed (see Fig.~15 in \citealt{hilton_atacama_2021}.) Therefore, an independent cross-match of the SNR peaks list with the HSC catalog is unlikely to identify many new ACT candidates as clusters. 

Unlike with optical data, SZ maps enable the identification of massive clusters at any redshift, as the strength of the SZ effect does not depend on the cluster's distance to the observer. For this reason, having a deep optical cluster catalog---in this case, the HSC catalog, which extends beyond $z \sim 1$---allows us to construct a uniquely large cross-matched sample. One does not expect all ACT clusters in the HSC footprint to have a match in the HSC catalog, because some ACT clusters may be at higher redshifts than those probed by HSC. (As mentioned in the previous paragraph, $\lesssim$ 5\% of ACT cluster candidates lack an optical counterpart.) On the other hand, one does not expect all HSC clusters to have a match in the ACT catalog: less massive HSC clusters are less likely to have high-pressure gas that gives rise to a measurable SZ signal.

\subsection{Fiducial Cross-Match}\label{subsec:cross_match_sample}

In creating the cross-matched sample for this study, we identify the nearest HSC cluster for each ACT cluster. We do so via the \texttt{astropy} package\footnote{Specifically, we use the method \texttt{match\_to\_catalog\_sky} from the \texttt{SkyCoord} class.} \citep{robitaille_astropy_2013, price-whelan_astropy_2018, astropy_collaboration_astropy_2022}. For a given cross-match between an ACT and HSC cluster, we calculate the (projected) physical offset by using the CAMIRA photometric redshift for the HSC cluster. We then impose a maximum possible physical offset of $1 \text{ Mpc/h}$ (i.e., a matching radius) for the pair to be considered a match. Because clusters are $\sim 1$ Mpc in size, a pair of HSC and ACT clusters separated by this distance or more are almost certainly distinct.

We only consider HSC clusters whose listed richness $\lambda$ is greater than 15. At very low richnesses, the HSC catalog is both less complete and more contaminated with false clusters due to projection effects. Furthermore, as discussed in Section~\ref{subsubsec:act_cat}, the ACT catalog is unlikely to contain low-mass/low-richness clusters, so cross-matches between an ACT signal and a low-richness HSC cluster would be suspect.

After imposing the matching radius, we remove cross-matches that include ACT clusters with an associated warning in the ACT catalog. There are six such cases: one ACT cluster has a warning due to a potentially unreliable redshift; three clusters are associated with systems that may be two clusters at different redshifts projected onto each other; and the last two clusters were only confirmed via ``scanning mode,'' i.e., by running CAMIRA on an SZ position and identifying an optical cluster there that was not originally in the HSC catalog. We remove these latter two clusters because their optical centers are dependent on ACT data.

Lastly, there is one case in which two ACT clusters are matched to the same HSC cluster. Here, we accept the lower-redshift ACT cluster as the true match. The resulting sample contains 186 clusters with redshifts between 0.1 and 1.4 within an area of $469 \text{ deg}^2$; this is the largest existing cross-matched sample that uses HSC data. Throughout this paper, we refer to this sample as the fiducial cross-match or the fiducial sample.


In addition to the fiducial sample, we create a cross-match where the matching radius is a function of ACT SNR. We do this using the \texttt{nemo} algorithm \citep{hilton_nemo_2023}, which also includes a conservative estimate for positional uncertainties in the optical catalogs by adding in quadrature an additional 0.5 Mpc projected distance to the cross-matching radius. This \texttt{nemo}-based sample excludes five clusters contained in the fiducial sample, but is otherwise identical. These five excluded clusters are all highly miscentered; thus, one should suspect that systematic effects in the cross-matching and/or the CAMIRA algorithm may be relevant to these five cases. \ACTrevision{Indeed, in our analysis described in Section~\ref{sec:causes_of_miscent}, we find that three of the excluded clusters are miscentered due to a star mask, one is miscentered due to an observational artifact, and one is miscentered due to a false ACT signal.}

\subsubsection{Fiducial Cross-Match Properties}\label{subsubsec:cm_props}

\begin{figure*}
    \centering
    \includegraphics[width=0.95\textwidth]{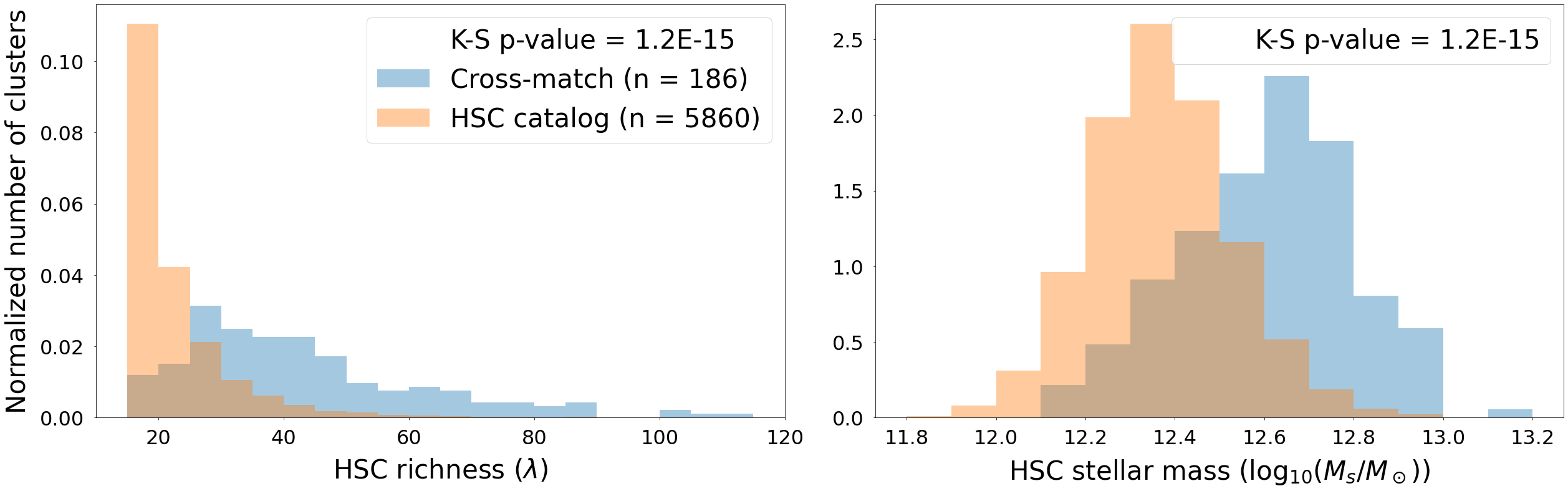}
     \caption[Properties of cross-match clusters vs. clusters in the full catalogs]{The normalized distributions of richness and stellar mass for both the fiducial cross-match and the full catalogs, along with the corresponding K-S test p-values. All p-values are statistically significant, indicating that the two catalogs are drawn from different underlying distributions. (We use SciPy's K-S function, whose minimum output p-value is effectively $1.2 \times 10^{-15}$.)
     }
     \label{fig:props_cm_full}
\end{figure*}

In Fig.~\ref{fig:props_cm_full}, we plot the distributions of richness $\lambda$ and stellar mass $M_s$ for both the fiducial cross-match and the full catalog. We use the values listed in the HSC catalog. In brief, HSC defines richness as the sum of the membership probabilities of all galaxies in a cluster, and stellar mass is defined as the weighted sum of the individual galaxy masses, each of which are calculated using a stellar population synthesis model \citep{oguri_cluster_2014}. In our richness distributions, the number of clusters generally decreases with richness, consistent with similar studies that use different optical and SZ catalogs (e.g., \citealt{grandis_exploring_2021}).

\begin{figure}
     \centering
\includegraphics[width=0.49\textwidth]{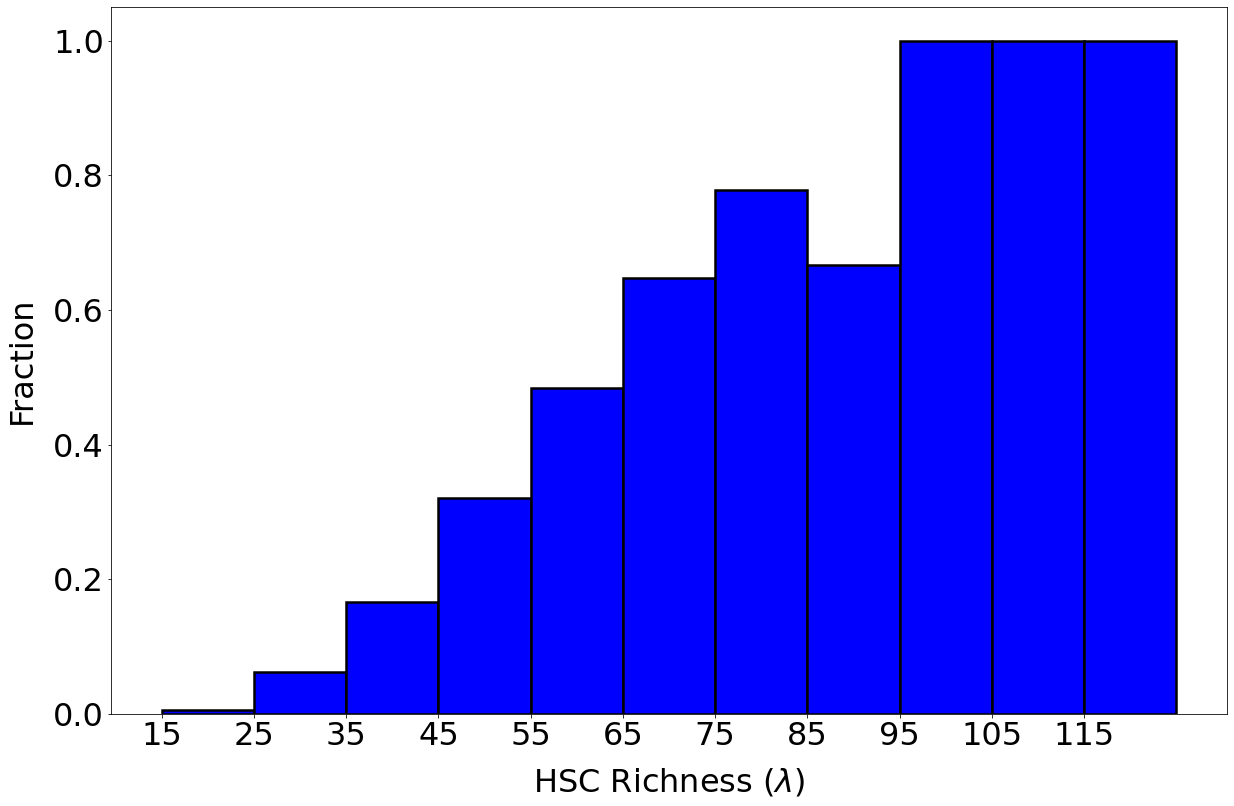}
     \caption[ACT completeness]{ACT completeness as a function of HSC richness. Completeness generally increases with richness, reaching 100\% for clusters with $\lambda \gtrsim 100$.}
     \label{fig:act_completeness}
\end{figure}

For both properties, we perform a two-sided Kolmogorov-Smirnov (K-S) test, which estimates the probability that two samples come from the same underlying continuous distribution; in the context of this work, we test if our cross-matched sample is representative of the underlying catalogs. As shown in Fig.~\ref{fig:props_cm_full}, the p-value for each property is statistically significant, suggesting that the fiducial cross-match and the full catalog have different underlying distributions. In particular, one sees that cross-match clusters are generally richer and have higher stellar mass. These trends are expected because, as Fig.~\ref{fig:act_completeness} shows, richer/more massive clusters generally have stronger SZ signals; thus, lower richness/lower mass HSC clusters are less likely to have a match in the ACT catalog. 



Fig.~\ref{fig:act_completeness} shows the completeness of the cross-matched sample as a function of HSC richness. This completeness is defined as the fraction of HSC clusters within a given richness bin that have a counterpart in the ACT catalog. Consistent with similar cross-matching studies (e.g., \citealt{grandis_exploring_2021}), the completeness increases with richness, reaching $1$ for the five clusters with $\lambda \gtrsim 100$.

The full sample of matched clusters, together with their offsets and our interpretations of them, are given in Appendix~\ref{sec:list_of_cm_clusters}.  

\section{Centering Offset Model}\label{sec:offset_model}

\subsection{Offset Distribution and Model}\label{subsec:offset_distro}

\begin{figure*}
     \centering
     \includegraphics[width=0.95\textwidth]{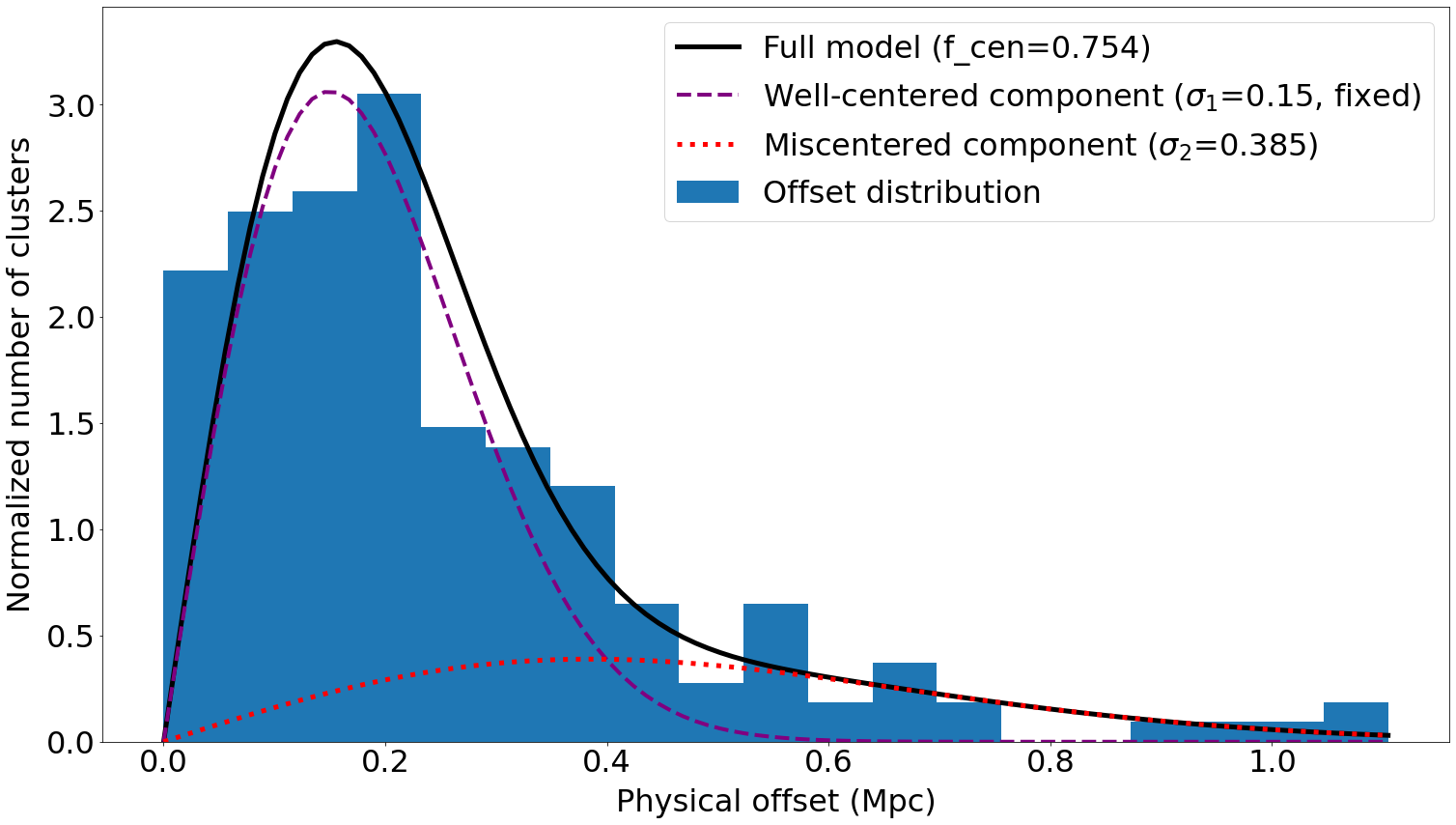}
     \caption[Offset distribution for the fiducial cross-match]{Offset distribution for the fiducial cross-match, including the histogram of measured offsets, the best-fit model using Equation~\ref{eq:distro_eq}, and the two components of the model.}
     \label{fig:fiducial_distro}
\end{figure*}

We aim to construct a model of the distribution of centering offsets $r$ within our fiducial cross-match. Here, $r$ is the projected physical separation at the redshift of the cluster. In doing so, our goal is to infer the fraction of clusters that have an offset due to measurement uncertainty and the fraction that are miscentered due to other effects, such as mergers of clusters.

The histogram in Fig.~\ref{fig:fiducial_distro} shows the normalized distribution of $r$ for our fiducial cross-match. We model this distribution using the two-component model from \cite{oguri_optically-selected_2018}:

\begin{equation}
    p(r)  = f_{\text{cen}} \frac{r}{\sigma_1^2} \text{ exp}\left(-\frac{r^2}{2\sigma_1^2} \right) + \left(1 - f_{\text{cen}}\right) \frac{r}{\sigma_2^2} \text{ exp}\left(-\frac{r^2}{2\sigma_2^2} \right) 
    \label{eq:distro_eq}
\end{equation}

The first component models the well-centered cluster population, while the second component models the miscentered population; each individual component is modeled with a Rayleigh distribution, which assumes that the horizontal and vertical displacements on the sky are each normally distributed. Here, $r$ is the offset in Mpc, $f_\text{cen}$ is the fraction of clusters that are well-centered, and $\sigma_1$ and $\sigma_2$ are the characteristic offsets for the well-centered and miscentered populations, respectively. 

We expect the positional error to be dominated by the SZ positional uncertainty: while optical astrometric errors on individual galaxies are less than 1 arcsec, the SZ positional uncertainty is on the order of $\gtrsim 10"$ due to the resolution of the CMB maps. Given a typical value of $\text{SNR}_{2.4} = 5$ for the fiducial sample, we expect a positional error of $\sim 17"$ (see \citealt{hilton_atacama_2021}, who run the \texttt{nemo} algorithm on simulated ACT data and model the distribution of SZ positional uncertainty as a function of $\text{SNR}_{2.4}$). This angle corresponds to a projected offset of $\sim 0.15$ Mpc over most of the redshift range of the catalog. We thus fix $\sigma_1 = 0.15$ Mpc, and  
fit for $f_\text{cen}$ and $\sigma_2$ simultaneously with a maximum likelihood method.

Fig.~\ref{fig:fiducial_distro} shows the measured and best-fit offset distribution of the fiducial cross-match. The best-fit parameters are \ACTrevision{$f_\text{cen} = 0.75$ and $\sigma_2 = 0.39 \text{ Mpc}$}. This $f_\text{cen}$ is consistent with previous miscentering studies, while $\sigma_2$ is consistent with \citep{oguri_optically-selected_2018}, which reports $\sigma_2 = 0.26 \pm 0.04 \, h^{-1}$ Mpc, approximately $\sigma_2 \sim 0.37$ Mpc for $h=0.7$.

\subsection{Well-Centered vs. Miscentered Populations}\label{subsec:well_vs_mis}

\begin{figure*}
     \centering
\includegraphics[width=0.95\textwidth]{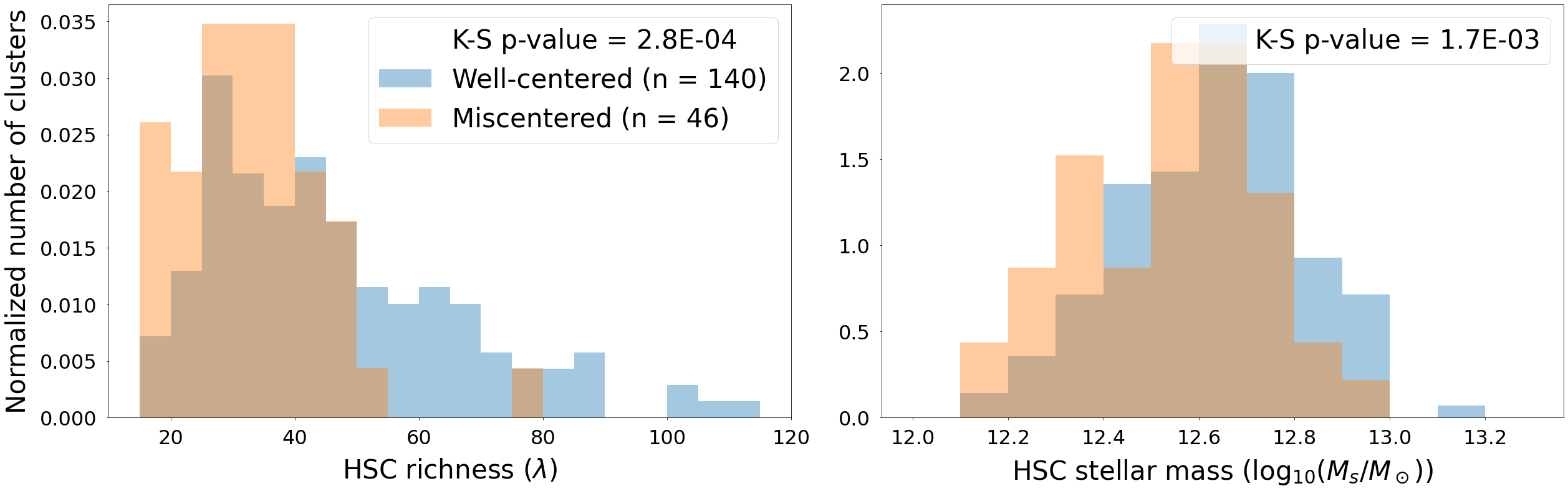}
     \caption[Properties of well-centered vs. miscentered clusters within the fiducial cross-match]{The normalized distributions of richness and stellar mass for both well-centered and miscentered clusters within the fiducial cross-match, along with the corresponding K-S test p-values. The p-values for $\lambda$ and $M_s$ are both statistically significant, indicating that miscentered clusters have lower richness and stellar mass than well-centered clusters. 
     }
     \label{fig:props_well_mis}
\end{figure*}

Based on our best-fit model, $f_\text{cen} = 0.754$ of the clusters in the fiducial cross-match are well-centered\ACTrevision{; these clusters have offsets of $r < 330$ kpc.} Thus, we denote this value, 330 kpc, as the \textit{well-centered cutoff}. The remainder of the clusters are labeled as miscentered. We acknowledge that this definition of the well-centered cutoff is somewhat arbitrary. In Section~\ref{sec:lensing_measurements}, we show that the difference in lensing signals between the well-centered and miscentered populations is statistically significant, thus validating our choice of the well-centered cutoff.

Following Fig.~\ref{fig:props_cm_full}, Fig.~\ref{fig:props_well_mis} compares the distributions of $\lambda$ and $M_s$ between the well-centered and miscentered clusters. Here, the p-values of the K-S test for $\lambda$ and $M_s$ are statistically significant, suggesting that well-centered and miscentered clusters are not drawn from the same underlying distribution. Miscentered clusters generally have lower $\lambda$ and $M_s$ than do well-centered clusters. This may partially be explained by the fact that the inferred richness and mass of miscentered clusters are biased low by miscentering.

\section{Causes of Miscentering}\label{sec:causes_of_miscent}

In this section, we discuss potential causes of miscentering in the fiducial sample. Using the HSC S19A wide-field data release \citep{aihara_third_2022}, we individually examine the optical images of all 46 miscentered clusters in the fiducial sample. The cluster images shown in this paper (e.g., Fig.~\ref{fig:example_merger}) are composites of images taken in the $g$, $r$, and $i$ filters.

We find that causes of miscentering broadly \ACTrevision{fall into} three categories: cluster mergers, systematic effects in the HSC data, and other causes such as potential false matches and false ACT signals. We also discuss miscentered clusters that seemingly lack any cause of miscentering. Table~\ref{tab:causes_of_misc} summarizes the eight causes of miscentering discussed in our analysis, while Table~\ref{tab:fiducial_cm} has the full list of clusters in our cross-match, including miscentering causes as relevant.

\begin{table}
\begin{tabular}{|c|c|}
\hline
\textbf{Cause of Miscentering} & \textbf{Number of Clusters} \\ \hline
Merger (\S\ref{subsec:mergers})                & 14            \\ \hline
Star mask   (\S\ref{subsubsec:star_masks})         & 11            \\ \hline
Observational artifact (\S\ref{subsubsec:artifacts}) & 2             \\ \hline
Deblending   (\S\ref{subsubsec:deblending})        & 2             \\ \hline
Misidentified central galaxy  (\S\ref{subsubsec:central})  & 2             \\ \hline
False match    (\S\ref{subsubsec:false_match})      & 2             \\ \hline
False ACT signal (\S\ref{subsubsec:false_ACT_signal})              & 3             \\ \hline
Multiple possible causes  (\S\ref{subsubsec:multiple})     & 6             \\ \hline
No apparent cause  (\S\ref{subsubsec:no_cause})            & 4             \\ \hline
\end{tabular}

\caption{A summary of the causes of miscentering identified in our analysis along with the number of miscentered clusters corresponding to each cause. We indicate the section of the text where each cause is described. The ``multiple possible causes'' category indicates clusters that may be affected by multiple different causes of miscentering, and the ``no apparent cause'' category indicates clusters without any apparent cause of miscentering.}
\label{tab:causes_of_misc}
\end{table}

\subsection{Merging Clusters}\label{subsec:mergers}

As discussed in Section~\ref{sec:intro}, cluster mergers are a potential astrophysical explanation for clusters with a large offset between the optical and SZ-defined centers. In merging systems, the gravitational potential may well not have a well-defined center. Even if this center is well-defined, then there might not be a bright, massive galaxy that has moved close enough to that center yet. To determine which of our miscentered clusters are due to ongoing mergers, we compare our fiducial cross-match with a pre-existing catalog of merging clusters in the HSC catalog. This merger catalog was created by looking for clusters with multiple peaks in their projected galaxy distributions \citep{okabe_halo_2019}. 

\begin{table}
\centering
\begin{tabular}{|c|c|c|c|c|}
\hline & \textbf{Merging} & \textbf{Not Merging} & \textbf{Total} & \textbf{Merging Fraction} \\ \hline
\textbf{Well-centered} & 68           & 72           & 140       & 49\%    \\ \hline
\textbf{Miscentered}   & 23           & 23            & 46        & 50\%    \\ \hline
\textbf{Total}         & 91           & 95           & 186       & 49\%    \\ \hline
\end{tabular}
\caption[Merging statistics for the fiducial cross-match]{Merging statistics for the fiducial cross-match, based on a catalog of optically-identified merging clusters in the HSC catalog \citep{okabe_halo_2019}.}
\label{tab:merg_stats}
\end{table}
Table~\ref{tab:merg_stats} shows the results of the comparison, which indicates that 50\% of the miscentered clusters are undergoing mergers. We emphasize, however, that not all clusters identified as merging by \cite{okabe_halo_2019} are miscentered solely due to an ongoing merger. For example, a cluster identified as a merger by \cite{okabe_halo_2019} could also be miscentered due to a star mask over the true central galaxy (see Section~\ref{subsubsec:star_masks}). Therefore, the number of merging, miscentered clusters as listed in Table~\ref{tab:merg_stats} exceeds the number of merging, miscentered clusters we identify in our sample (see Table~\ref{tab:causes_of_misc}).

The fraction of miscentered clusters which are merging is similar to that for well-centered clusters, indicating that mergers are \textit{not} strongly correlated with miscentering. Furthermore, a non-negligible fraction of miscentered clusters are not mergers. It is possible that the merger catalog in \cite{okabe_halo_2019} is incomplete (i.e., missing some merger systems) or that there are other, not necessarily astrophysical processes that also contribute to miscentering. Investigating the former is outside the scope of this paper, as it would require an independent probe of cluster mergers; however, we explore the latter possibility in detail in Section~\ref{subsec:hsc_effects}. 

\begin{figure}
\centering
\includegraphics[width=0.47\textwidth]{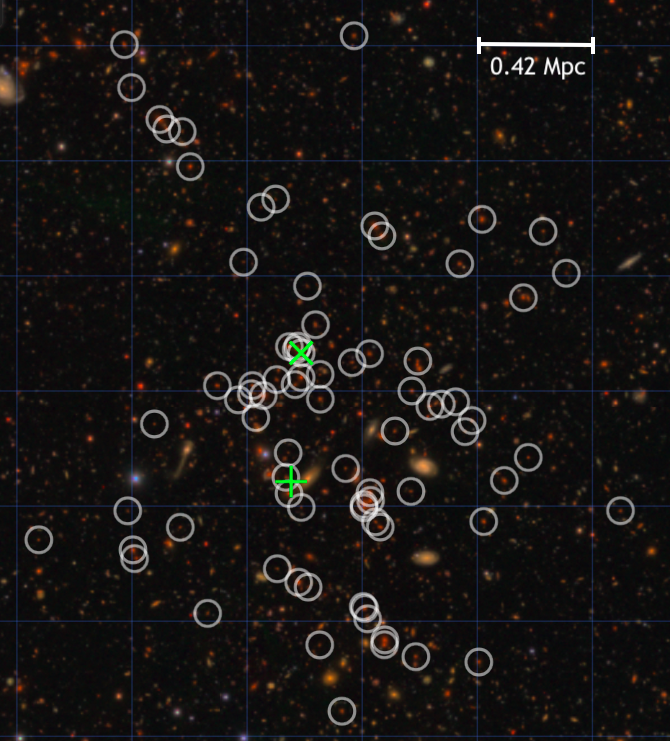}
     \caption{HSC cluster HSCJ142103+002322 at redshift $z = 0.65$, an example of a miscentered cluster that can be attributed to a merger. It has an offset of 0.47 Mpc. The green $+$ symbol indicates the ACT center, the green $\times$ symbol indicates the HSC center, and the white circles indicate member galaxies of the cluster with membership probability $>0.1$ (we use this convention in subsequent figures as well). In this and subsequent figures, North is up and East is to the right; note that this differs from the usual convention of having East be to the left.
     The scale is shown with the scale bar in the figure. There appear to be three potential peaks in the cluster's galaxy distribution: one at the HSC center, one to the right of the ACT center, and one near the bottom of the image.}
     \label{fig:example_merger}
\end{figure}

In Fig.~\ref{fig:example_merger}, we show an example of a cluster that we determine to be miscentered due an ongoing merger. In agreement with \cite{okabe_halo_2019}, we identify this cluster as a merger because there appears to be three potential peaks in this cluster's galaxy distribution. Additionally, the ACT center is located between these peaks, further suggesting that this system consists of merging clusters.

In Appendix~\ref{sec:props_merg_vs_non}, we compare the properties of merging and non-merging clusters within the cross-match.

\subsection{Systematic Effects in HSC Data}\label{subsec:hsc_effects}

Within the category of systematic HSC effects, we identify four causes of miscentering: masks due to bright stars, observational artifacts, galaxy deblending issues, and CAMIRA misidentification of the cluster's central galaxy. In total, 17 miscentered clusters \ACTrevision{fall into} this category.

\subsubsection{Star masks}\label{subsubsec:star_masks}

\begin{figure}
     \centering
     \includegraphics[width=0.45\textwidth]{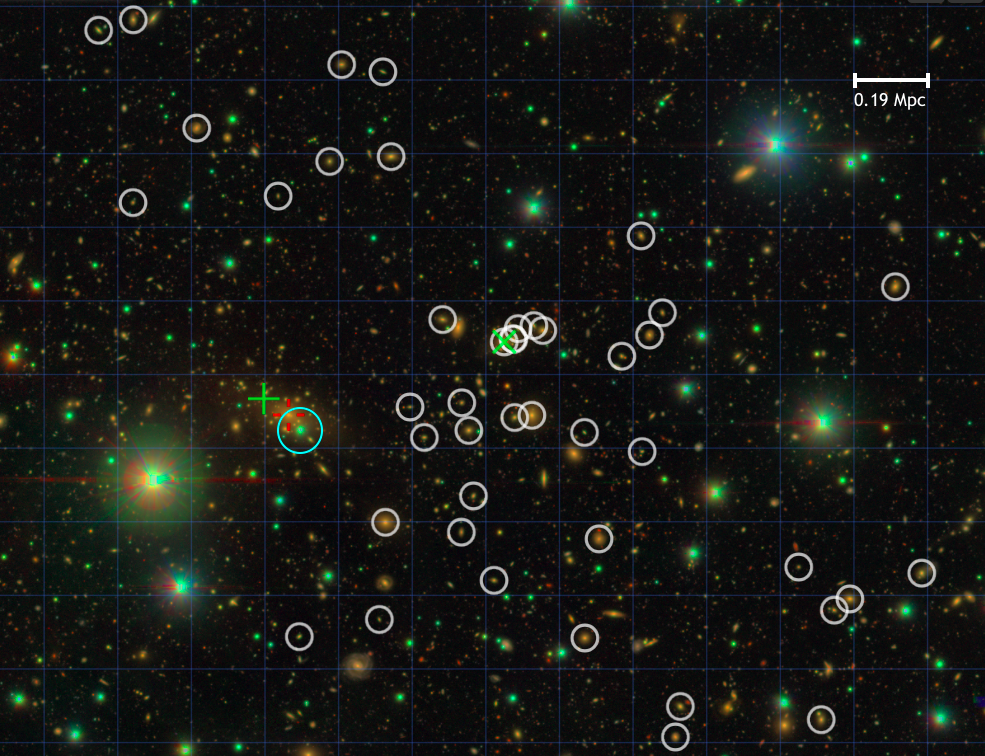}
     \caption{HSC cluster HSCJ090119+030156 at $z=0.19$, an example of a miscentered cluster that can be attributed to the obscuring effects of a bright star with $G_\text{Gaia} \sim 14.7$; this star is named Gaia DR3 578211316549315584 \citep{gaia_collaboration_vizier_2022}. The cluster has an offset of 0.65 Mpc. The red crosshairs (near the ACT center) indicate the alternative central galaxy that is obscured by the star's circular mask, represented by the cyan circle. The offset using this alternative central galaxy is only 0.08 Mpc. }
     \label{fig:example_star}
\end{figure}

Like all optical imaging surveys, HSC is affected by bright stars, whose extended point spread function can affect the detection and photometry of galaxies in their vicinity. For this reason, the galaxy sample on which the CAMIRA cluster catalog is based has been trimmed of objects falling too close to bright stars \citep{coupon_bright-star_2018}. However, if the resulting star mask overlaps the footprint of a cluster, the determination of its center is likely to be biased or completely incorrect.

We find 11 clusters that are most likely miscentered due to the star mask within the HSC data. In these cases, we identify a potential alternative central galaxy at the same photometric redshift as the cluster, with high luminosity and an extended envelope. These alternative central galaxies were not identified by CAMIRA since they are either fully or partially covered by an HSC star mask; instead, CAMIRA has chosen a different central galaxy, one that is not representative of the cluster center.

We determine if a star mask is obscuring an alternative central galaxy by calculating the extent of its star mask. \cite{coupon_bright-star_2018} describe the procedure for building the HSC star masks: a major component for each mask is a circular mask with radius depending on the star's $G$-band magnitude in the Gaia catalog \citep{gaia_collaboration_gaia_2023}:
\begin{equation}
    r [\text{arcsec}] =
        \begin{cases}
        708.9 \times \text{exp}(-G_\text{Gaia}/8.41), G_\text{Gaia} < 9 \\
        694.7 \times \text{exp}(-G_\text{Gaia}/4.04), G_\text{Gaia} \geq 9. \\
        \end{cases}
\label{eq:mask_radius}
\end{equation}

In Fig.~\ref{fig:example_star}, we show an example of such a cluster, with an apparent offset of 0.65 Mpc. The large elliptical galaxy with an extended envelope near the ACT center has a spectroscopic redshift of 0.19 \citep{ahn_ninth_2012}, the same as the cluster's photometric redshift, yet it is excluded from the cluster's member list. Furthermore, the offset between the ACT center and the alternative central galaxy is only 0.08 Mpc. The exclusion of this galaxy from the CAMIRA member galaxy list is a result of the mask around the bright star with radius $18''$. Because of this excluded galaxy, we believe that the CAMIRA-determined center is likely incorrect, causing this cluster to be labeled as miscentered.

\subsubsection{Observational artifacts}\label{subsubsec:artifacts}

Similar to the star mask case, observational artifacts can contaminate images of HSC clusters, preventing CAMIRA from detecting the true central galaxy of a cluster. We find two such cases where an observational artifact either impacts the imaging of the entire cluster or passes through a likely central galaxy, resulting in that central galaxy not being identified by CAMIRA. Both of these cases appear to be the result of improper stacking of single-exposure images, resulting in bright streaks on the co-added image.

\subsubsection{Deblending issues}\label{subsubsec:deblending}

\begin{figure*}
     \centering
     \includegraphics[width=0.45\textwidth]{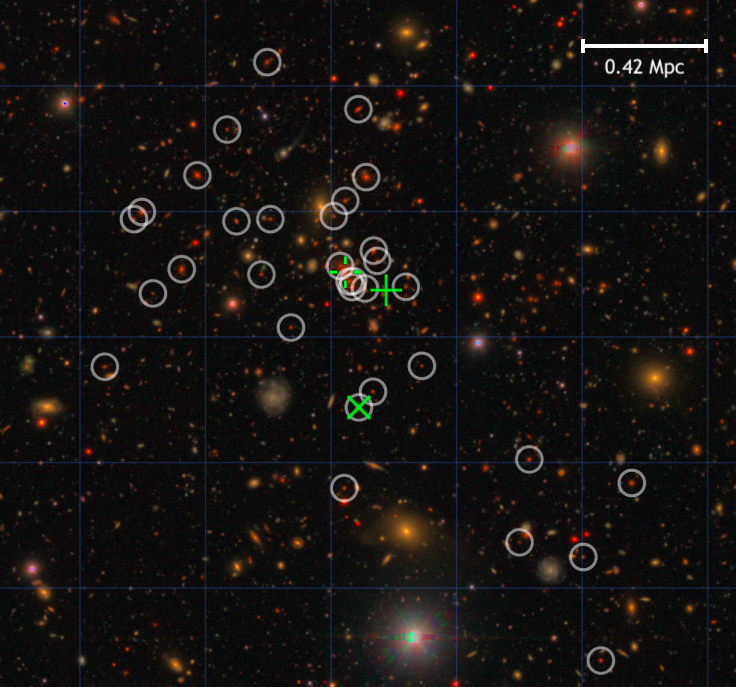}
    \includegraphics[width=0.45\textwidth]{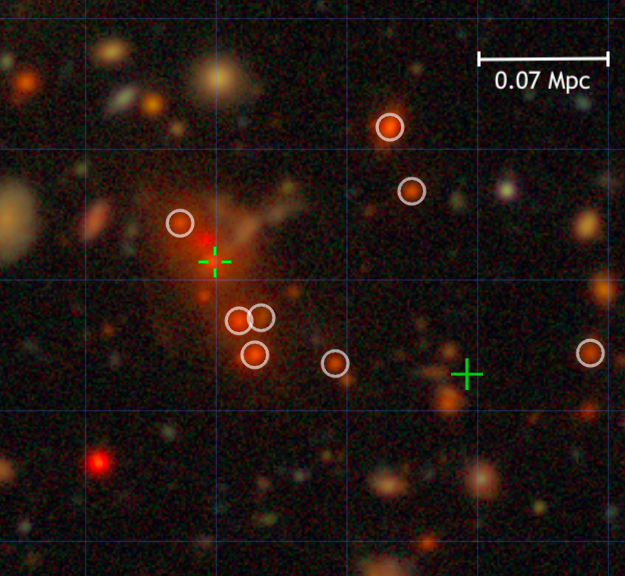}
     
     \caption{\textbf{Left:} HSC cluster HSCJ021002-024411 at $z \approx 0.66$, an example of a miscentered cluster that can be attributed to HSC image deblending issues. It has offset 0.40 Mpc. The green crosshairs indicate a potential alternative galaxy. 
     \textbf{Right:} A close-up view of the same cluster near the ACT center. Many of the overlapping galaxies near the ACT center are not identified as individual sources in the HSC object catalog, suggesting that they are not properly deblended. It is unclear if the marked object is a single galaxy, two galaxies in the process of merging, or two physically distinct galaxies; further, it is unclear whether one of these galaxies could be the true central galaxy of the cluster. 
     }
     
     \label{fig:example_deblending}
\end{figure*}

In the crowded environments of galaxy clusters, the images of galaxies often overlap and need to be deblended to measure the brightness of each galaxy separately. The deblending algorithm, described in \cite{bosch_hyper_2018}, occasionally fails, giving incorrect photometry of the individual galaxies. We found two such cases among our sample of miscentered clusters. 

These deblending failures are identifiable by the fact that not every visible galaxy or component of a blend was identified in the HSC object catalog. If deblending fails, then the true central galaxy may have an incorrect reported brightness or not be detected at all, among other potential issues. In either case, CAMIRA would select another galaxy as the center, leading the cluster to be labeled as miscentered. We find two such miscentered clusters, one of which is shown in Fig.~\ref{fig:example_deblending}.

\subsubsection{Central galaxy misidentification}\label{subsubsec:central}

For two miscentered clusters, we determine that the potentially true central galaxy was detected by CAMIRA and labeled as a member of the cluster---however, CAMIRA still chose a different galaxy as the center. In particular, we see a large, elliptical galaxy in the cluster that is not chosen as the center despite being both brighter and closer to the ACT center than the CAMIRA-chosen center. For both clusters, if the alternative central galaxy were selected instead, then the cluster's offset would fall below the well-centered cutoff of 330 kpc (Section~\ref{subsec:well_vs_mis}).

\subsection{Other Causes}\label{subsec:other_causes}

We identify two causes of miscentering that do not fall in the category of either mergers or systematic HSC effects: false matches between ACT and HSC clusters, and false ACT signals. We also identify two other subcategories: clusters that could be miscentered due to multiple potential factors, and clusters that lack any apparent cause of miscentering. In total, 15 miscentered clusters fall into the category of other causes.

\subsubsection{False match}\label{subsubsec:false_match}

\begin{figure*}
     \centering
     \includegraphics[width=0.47\textwidth]{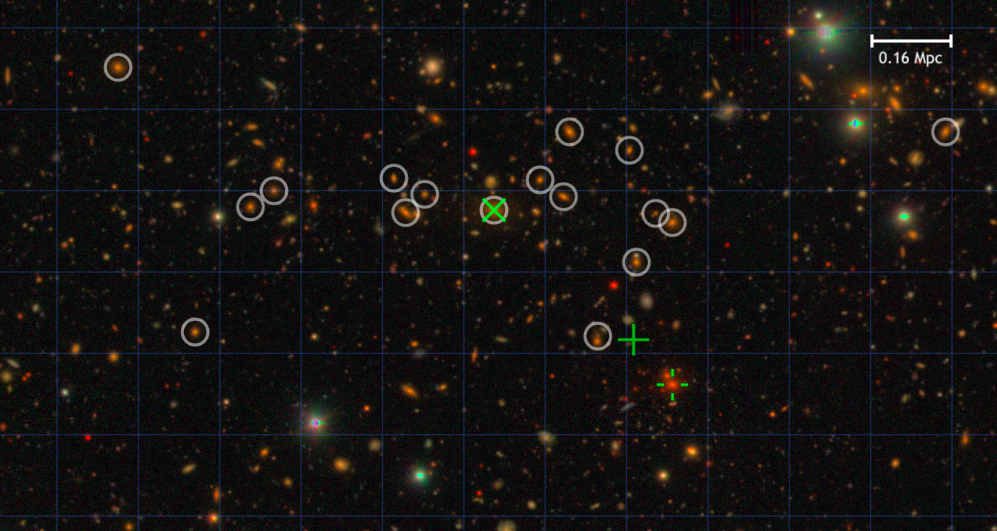}
     \includegraphics[width=0.45\textwidth]{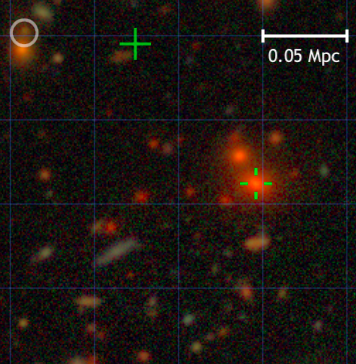}

     \caption{\textbf{Left:} HSC cluster HSCJ114409+044133 at $z \approx 0.41$, an example of a miscentered cluster that can be attributed to a false match between this cluster and an ACT cluster. It has offset 0.38 Mpc. We suspect that a galaxy overdensity near the ACT signal is an optical cluster missed by CAMIRA; its possible center is indicated with the green crosshairs. If this cluster were the true source of the detected SZ signal, then its offset would only be 0.14 Mpc. 
     \textbf{Right:} A close-up view of HSC cluster HSCJ114409+044133 near the center of the potential second cluster. 
     }
     
     \label{fig:example_falseMatch}
\end{figure*}

We suspect that two of the miscentered clusters arise from a false match between HSC and ACT clusters; i.e., when the corresponding ACT cluster is more likely to correspond to a different optical cluster that is not detected by CAMIRA. In both cases, the HSC cluster in the fiducial cross-match has a low richness ($ < 17 $). As discussed in Section~\ref{subsec:cross_match_sample}, low-richness clusters tend to have weak SZ signals and are often undetected by ACT, making the match less convincing, and strengthening our conclusion that these clusters are miscentered due to false matches.

One example cluster is shown in Fig.~\ref{fig:example_falseMatch}. The ACT center is $\sim 21''$ away from an overdensity of galaxies at $z\sim 0.6$, which we suspect is an optical cluster unidentified by CAMIRA. If this is the true optical cluster corresponding to the ACT center, then the offset would only be $\sim 0.14$ Mpc, which would change the cluster from miscentered to well-centered.

\subsubsection{False ACT signal}\label{subsubsec:false_ACT_signal}

\begin{figure}
     \centering
     \includegraphics[width=0.45\textwidth]{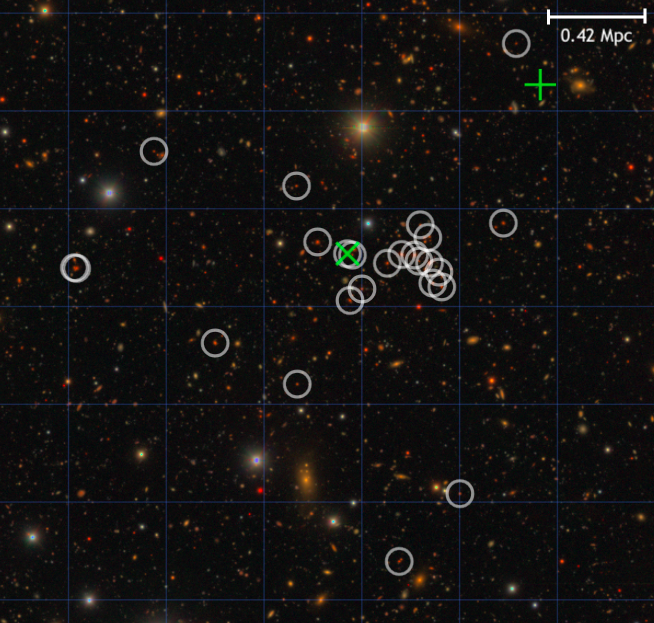}
     
     \caption{HSC cluster HSCJ135746+002431 at $z \approx 0.67$, an example of a miscentered cluster that can be attributed to this cluster being matched with a false ACT signal. There is no apparent galaxy overdensity near the ACT center. The cluster has offset 1.1047 Mpc. 
     }
     \label{fig:example_falseSignal}
\end{figure}

Based on Fig.~6 in \cite{hilton_atacama_2021}, one should expect around 2000 false ACT signals---i.e., a signal not originating from a real cluster---within the full ACT DR5 map. These false signals are due to noise fluctuations in the CMB maps. It would be reasonable to assume that our cross-match contains at least a few spurious matches between a false ACT signal and an HSC cluster. Additionally, since matches between false signals and HSC clusters are essentially random, it is reasonable to expect occasional matches with large offsets that nevertheless fall within our matching radius of $1 \text{ Mpc/h}$. Assuming these signals are randomly distributed, then we expect $\sim 70$ false signals in the HSC footprint and estimate that $\sim 13$ of them would be cross-matched with an HSC cluster \ACTrevision{given that our matching circles cover $\sim$20\% of the ACT-HSC overlap region}. As mentioned in Sec.~\ref{subsubsec:act_cat}, \cite{hilton_atacama_2021} finds that more than 95\% of ACT cluster candidates within the HSC footprint are optically confirmed. However, given our calculation, it is feasible that at least a few of the real optical clusters (especially low-richness ones) were matched to false ACT signals. 

In our sample, we find three miscentered clusters that appear to correspond to false signals in the ACT catalog. For each cluster, we did not observe any apparent galaxy overdensity near the ACT center. We confirmed this by looking at the SZ signal's contours (as explained in Fig. 10 in \citealt{hilton_atacama_2021}), which had little to no overlap with the HSC cluster galaxy distribution. One such cluster is shown in Fig.~\ref{fig:example_falseSignal}. Because the ACT signal in the figure does not appear to correspond to any real cluster, we conclude that it is a false signal.

\subsubsection{Multiple possible causes}\label{subsubsec:multiple}

\begin{figure}
     \centering
     \includegraphics[width=0.45\textwidth]{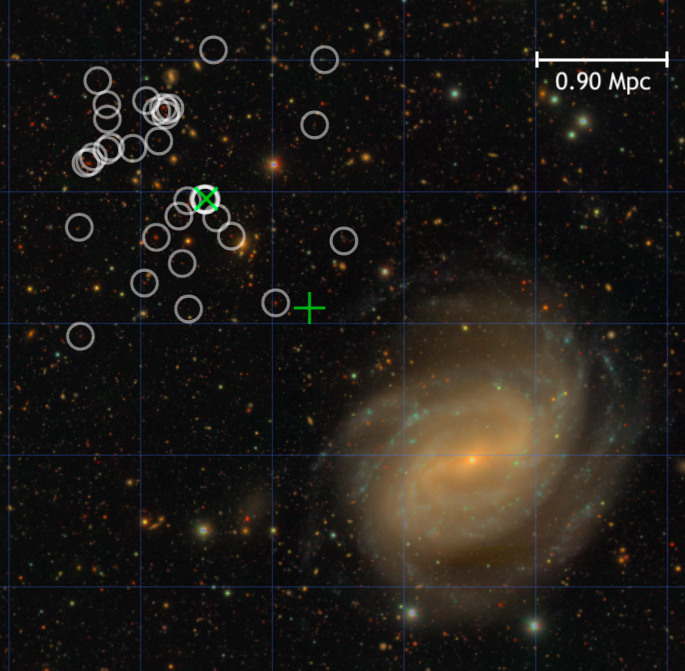}
     \caption{HSC cluster HSCJ120827+025640 at $z \approx 0.79$, an example of a cluster with multiple possible causes of miscentering: the obscuring foreground galaxy (NGC 4123), a merger, and/or a false ACT signal. It has offset 1.02 Mpc. 
      }
     \label{fig:example_multiple_possible}
\end{figure}

We identify six miscentered clusters that each appear to show symptoms of several causes of miscentering, no one of which was clearly the sole reason for the miscentering. Fig.~\ref{fig:example_multiple_possible} shows one such cluster. It is unclear if the nearby, foreground galaxy is obscuring any cluster galaxies that would otherwise be detected by CAMIRA. For example, the true central galaxy might be obscured yet close to the ACT center, which would mean that this cluster is actually well-centered. It is also possible that the cluster is undergoing a merger but the foreground galaxy is obscuring one half of the merging cluster---thus, we only see the ACT signal located between the two halves of the merger. On the other hand, it may be true that the foreground galaxy does not obscure any cluster galaxies. In this case, the ACT signal might be a false signal that aligned with this HSC cluster by chance. The cluster has a relatively low richness of $\sim 20$, meaning that it is unlikely to have a detectable ACT signal (although the obscuration of the bright galaxy could be causing CAMIRA to underestimate the richness). 

\subsubsection{No apparent cause of miscentering}\label{subsubsec:no_cause}

We identify four clusters that lack any apparent cause of miscentering. In contrast with the false signal clusters, these clusters each have an ACT signal near a galaxy overdensity. Therefore, in each case, we believe that the offset is astrophysically real despite there being no evidence for an ongoing merger. In fact, when we refit the offset distribution with a ``cleaned'' cluster sample (Section~\ref{subsec:cleaned_distro}), the subsequent model yields a higher well-centered cutoff---as a result, the model labels three out of four of the ``no apparent cause'' clusters as well-centered instead of miscentered.

\subsection{Cleaned Offset Distribution}\label{subsec:cleaned_distro}

\begin{figure*}
     \centering
     \includegraphics[width=0.95\textwidth]{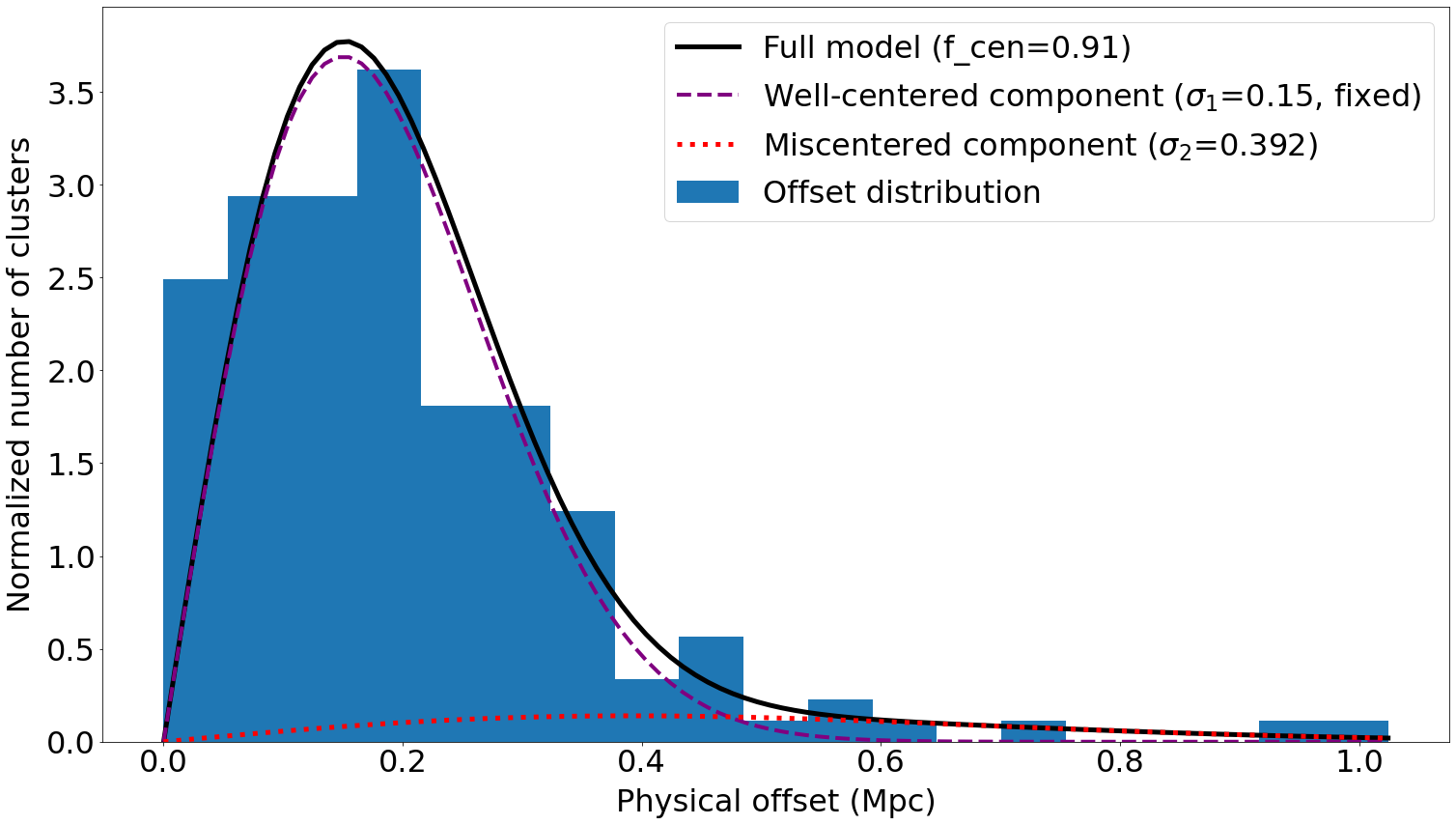}
     \caption{Offset distribution for the cleaned cross-match. The figure shows the offset histogram, the best-fit model using Equation~\ref{eq:distro_eq}, and the two components of the model, namely the well-centered and miscentered clusters.}
     \label{fig:cleaned_distro}
\end{figure*}

We create a ``cleaned'' cluster sample by taking the fiducial sample and excluding all miscentered clusters that have a clear, non-astrophysical cause of miscentering; i.e., the clusters that are miscentered due to a systematic HSC effect (Section~\ref{subsec:hsc_effects}), a false match (Section~\ref{subsubsec:false_match}), or a false ACT signal (Section~\ref{subsubsec:false_ACT_signal}). The resulting cleaned sample has 164 clusters. Fig.~\ref{fig:cleaned_distro} shows the offset distribution of this cleaned sample, together with the best-fit model (Equation~\ref{eq:distro_eq}). Here, we find $f_\text{cen} = 0.91$ and $\sigma_2 = 0.39$ Mpc.

This cleaned sample is separated into well-centered and miscentered populations using the same method described in Section~\ref{subsec:well_vs_mis}; i.e., the cutoff is chosen such that $f_\text{cen} = 0.91$ of the clusters are below the cutoff, and these clusters are labeled as well-centered. The cleaned sample's well-centered cutoff is \ACTrevision{370} kpc, slightly larger than the fiducial sample's well-centered cutoff of \ACTrevision{330} kpc. We find nine clusters within the cleaned sample that are miscentered in the fiducial sample but are well-centered in the cleaned sample; i.e., they each have an offset between \ACTrevision{330} and \ACTrevision{370} kpc. Six of these clusters were previously identified as being miscentered due to an ongoing merger, whereas the other three are ``no apparent cause'' clusters. (There are four ``no apparent cause'' clusters in total; the fourth, HSCJ231255-001521, has offset 535 kpc.) The fact that only one out of fifteen miscentered clusters in the cleaned sample is labeled as ``no apparent cause'' affirms that our offset model accurately separates our clusters into well-centered and miscentered populations.

\section{Lensing Measurements}\label{sec:lensing_measurements}

\begin{figure*}
     \centering
     \includegraphics[width=0.45\textwidth]{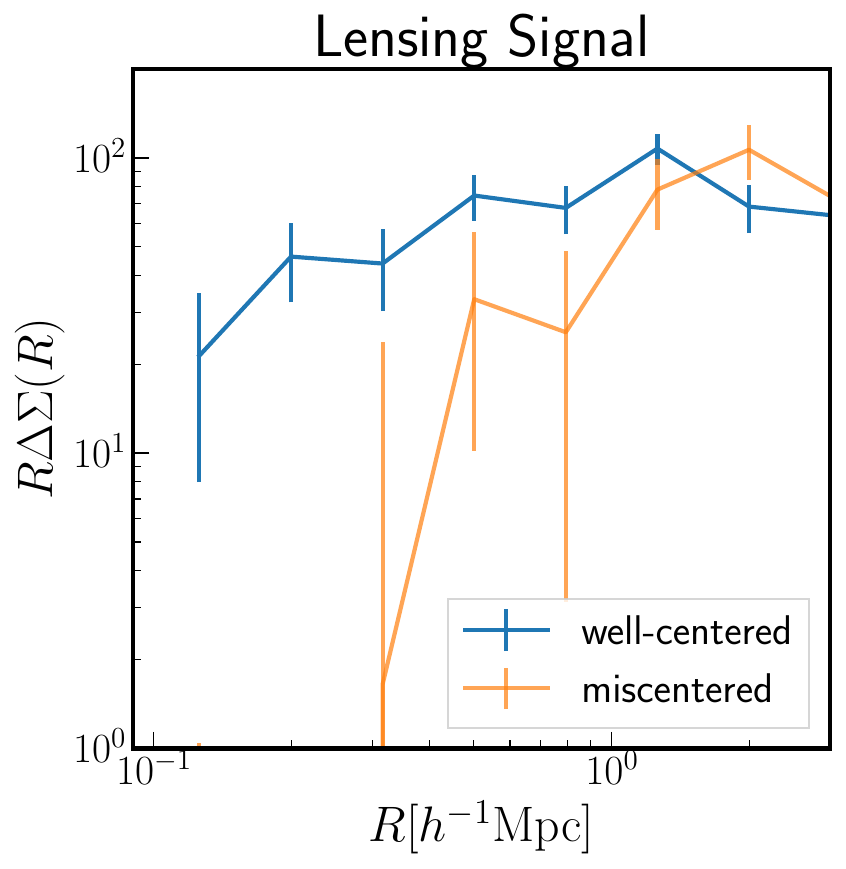} \hfill \includegraphics[width=0.45\textwidth]{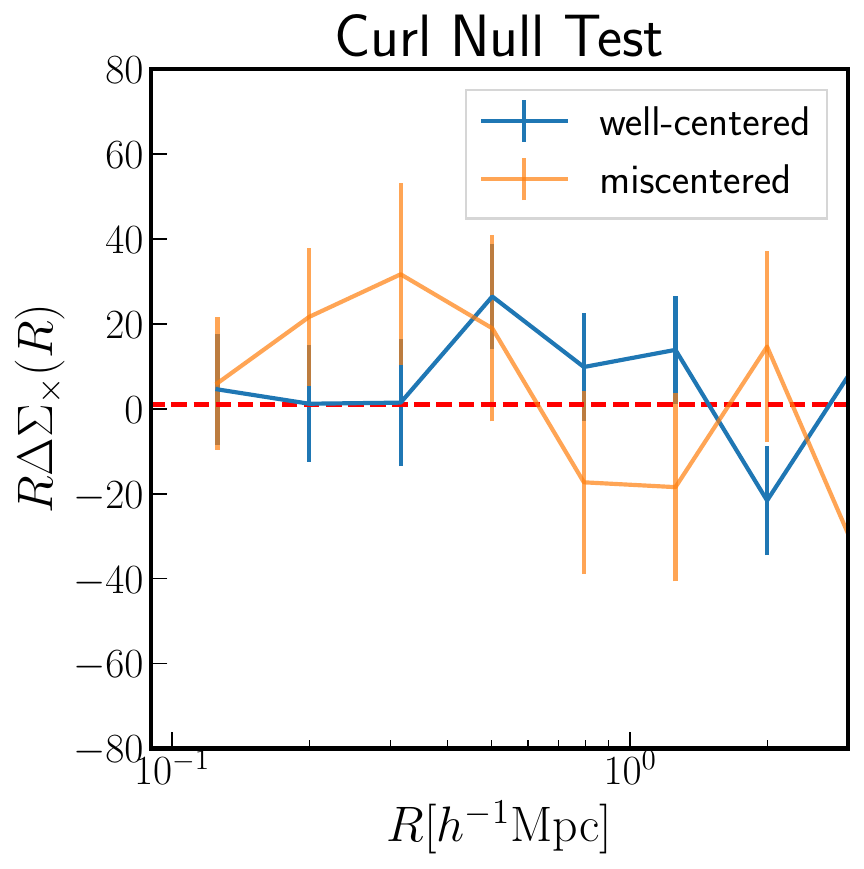}
     \caption{A comparison of the lensing signals for well-centered and miscentered clusters in the redshift range $0.3 < z < 0.7$. Here, we use the CAMIRA central galaxy as the cluster center. $\Delta \Sigma (R)$ is the cluster's lensing signal as a function of radius, and $\Delta \Sigma_\times (R)$ is the lensing signal in the 45-degree rotated component. The error bars represent the intrinsic shape noise. Note that $\Delta \Sigma$ is multiplied by $R$ to reduce the dynamic range in these plots. For the miscentered population, $\Delta \Sigma (R)$ is suppressed at small scales, as expected for miscentered clusters. $\Delta \Sigma (R)$ at the lowest radius is not shown for the miscentered clusters because it is negative with a large error bar. The difference in $\Delta \Sigma (R)$ between the well-centered and miscentered clusters is highly statistically significant, with a p-value of $6.6 \times 10^{-6}$.}
     \label{fig:lensing_well_vs_mis}
\end{figure*}

\begin{figure*}
     \centering
     \includegraphics[width=0.45\textwidth]{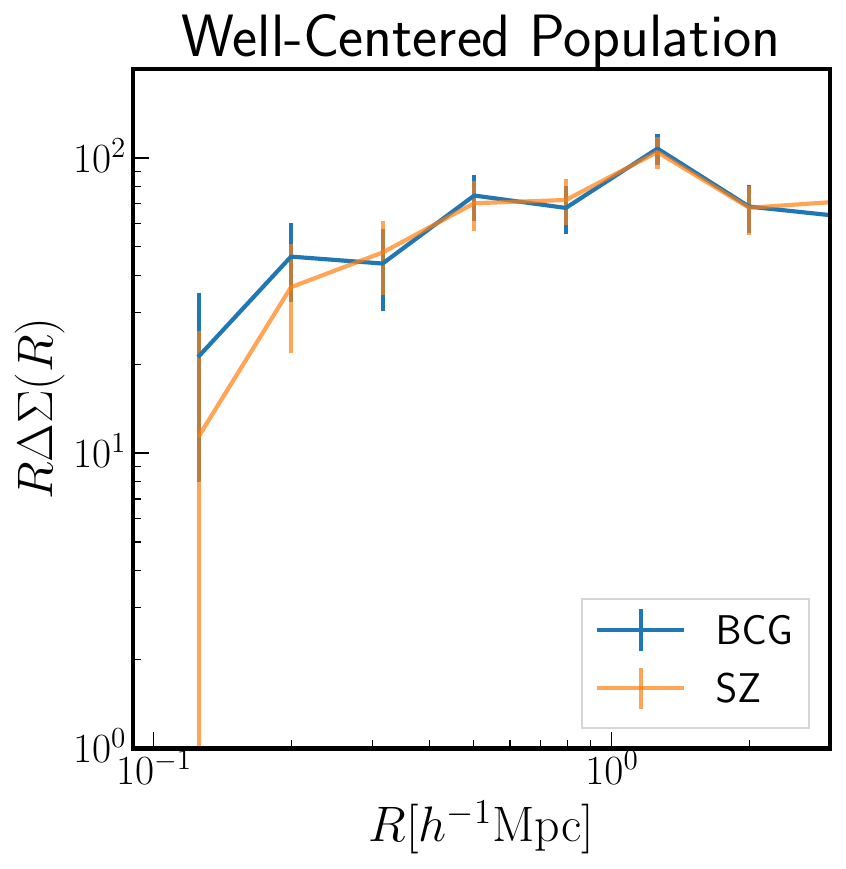} \hfill \includegraphics[width=0.45\textwidth]{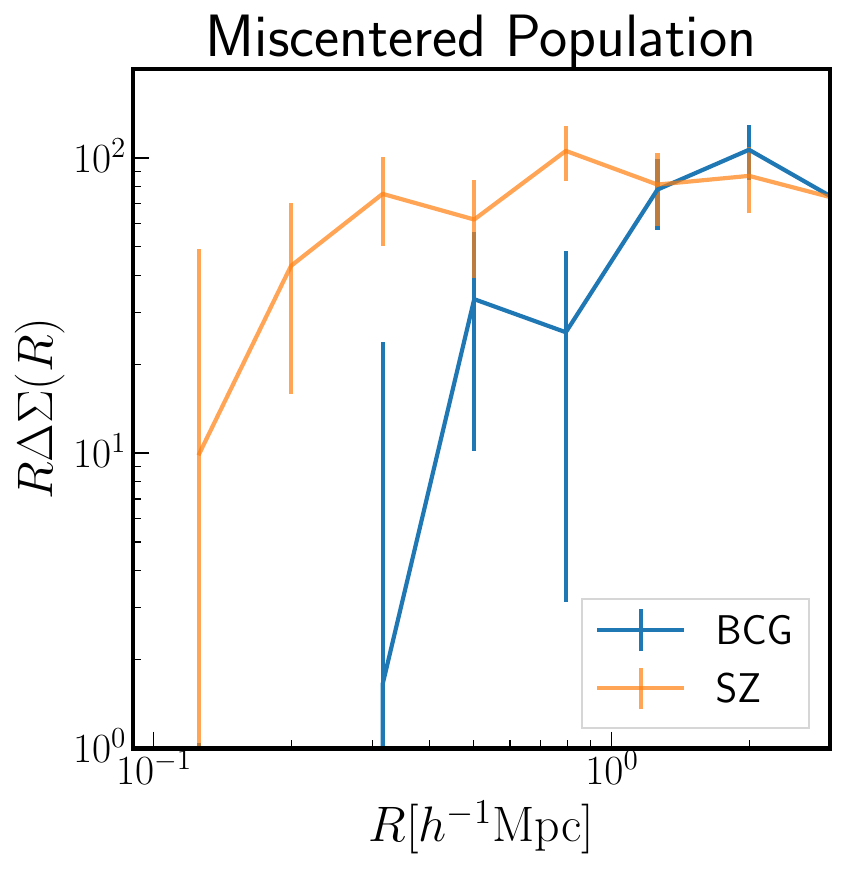}
     \caption{The measured lensing signal $R\Delta \Sigma (R)$ for both well-centered (left) and miscentered (right) clusters using the CAMIRA central galaxy (BCG) as the cluster center compared to using the ACT SZ center. These plots use clusters in the redshift range $0.3 < z < 0.7$. For the well-centered population, $\Delta \Sigma (R)$ is consistent regardless of whether the SZ or BCG center was used in the measurement. For the miscentered population, $\Delta \Sigma (R)$ is significantly suppressed at small scales when using the BCG center instead of the SZ center. $\Delta \Sigma (R)$ at the lowest radius is not shown for the miscentered clusters because it is negative with a large error bar. The difference in $\Delta \Sigma (R)$ between the two choices of centers is statistically significant for the miscentered population ($p=0.0276$) but is not statistically significant for the well-centered population.}
     \label{fig:lensing_bcg_vs_sz}
\end{figure*}

To understand the physical nature of the clusters we have identified as miscentered, we compare the weak gravitational lensing signals measured for the two sets of clusters using the Year 3 shape catalog of the HSC survey \citep{li_three-year_2022}. We follow the methodology outlined in \cite{sunayama_optical_2023} to measure the cluster lensing signal, with two main differences. Firstly, we select source galaxies with a redshift above $z_{l,\text{ max}}+0.2$ where $z_{l,\text{ max}}$ is the maximum redshift of the lenses (i.e., the clusters). We only consider clusters in the redshift range $0.3 < z < 0.7$ as we do not have a large enough sample of clusters or sources to make this measurement at higher redshifts. \ACTrevision{Within this range, there are 24 miscentered clusters and 72 well-centered clusters.} Secondly, we do not include boost and random corrections to the lensing signal. The former corrects for a dilution of the lensing signal due to source galaxies that are physically associated with the lens cluster, and the latter accounts for possible residual systematics in the shape measurement due to imperfect corrections of optical distortion across the field of view of the camera. Given the small sample size of our clusters, the measurement of the corrections is very noisy. Moreover, this paper focuses only on a comparison of lensing signals to identify potential suppression in the signal due to miscentering. Since there is no reason to expect that the corrections will be different for the populations of clusters that we compare (e.g. well-centered vs. miscentered), and we are not presenting the measurement of an absolute value of the signal, the corrections are unnecessary for the purposes of this work. For the same reason, we only use the shape-noise covariance matrix, which does not have any covariances between different angular bins. 

In Figure~\ref{fig:lensing_well_vs_mis}, we compare the lensing signals for the well-centered and miscentered clusters (as identified in Section~\ref{subsec:well_vs_mis}) in the redshift range $0.3 < z < 0.7$, using the CAMIRA central galaxy as the cluster center. $\Delta \Sigma  (R)$ is the cluster's lensing signal as a function of radius, and $\Delta \Sigma_\times (R)$ is the lensing signal in the 45-degree rotated component. The error bars represent the intrinsic shape noise. For the range of scales that we are interested in, and given the small sample size of our clusters, the shape noise is the dominant source of errors.
A non-zero $\Delta \Sigma_\times$ would suggest that there are systematic effects in the lensing signal that must be corrected, but for both the well-centered and miscentered clusters, our measurement of $\Delta \Sigma_\times$ is consistent with zero. 
The lensing signals of the two samples, shown in the left panel of Figure~\ref{fig:lensing_well_vs_mis}, are consistent on large scales ($>2 h^{-1}$ Mpc). However, the miscentered clusters show a suppressed signal below $\sim1$ $h^{-1}$ Mpc relative to the well-centered sample. This is what we would expect if the centroids of those clusters are systematically wrong; i.e., if the clusters are indeed miscentered. In fact, the difference in $\Delta \Sigma (R)$ between the well-centered and miscentered samples is statistically significant --- performing a $\chi^2$ test, we find $\chi^2=36.6$ with 10 degrees of freedom, giving a p-value of $6.6 \times 10^{-6}$. While one could in principle explore how this p-value varies as a function of the well-centered cutoff (which we defined in Section~\ref{subsec:well_vs_mis}), our results confirm that our selection of the two cluster populations is at least approximately accurate.  

We further estimate the impact of miscentering on the lensing signal by comparing the measured lensing signal for both well-centered and miscentered clusters using the CAMIRA central galaxy as the cluster center, relative to using the ACT SZ center. These results are shown in Figure~\ref{fig:lensing_bcg_vs_sz} for the well-centered (left panel) and miscentered clusters (right panel) in the redshift range $0.3 < z < 0.7$. For the well-centered population, whether we use BCG or SZ centers to measure $\Delta \Sigma (R)$, we get relatively consistent results. For the SZ centers, however, we see some suppression relative to the BCG center measurements at the smallest scales measured. Although this suppression is statistically insignificant \ACTrevision{(we find $\chi^2=0.98$ with 10 degrees of freedom, giving a p-value of 0.999843)}, we believe it could stem from the $\sim$0.15 Mpc positional uncertainty on the SZ center, as discussed in Section~\ref{subsec:offset_distro}. 

For the miscentered population, we see that the BCG and SZ center lensing signals are inconsistent with each other. The lensing signals measured using the BCG center are suppressed at small scales relative to those measured with the SZ centers. \ACTrevision{The difference in $\Delta \Sigma (R)$ between the two signals is statistically significant --- we find $\chi^2=20.18$ with 10 degrees of freedom, giving a p-value of 0.027595.} This suggests the SZ centers are more likely to represent the true cluster center than the BCG centers are. This verifies our intuition that, for miscentered clusters, one or more mechanisms cause CAMIRA to place the center far away from the true BCG, or clusters in the process of merging have BCGs far from the merged cluster center; in either case, the signal is suppressed. This makes the SZ center (i.e., the bottom of the gas potential well) more reliable for making lensing measurements.

\section{Discussion and Conclusions}\label{sec:discussion}

Utilizing the CAMIRA and ACT cluster catalogs---based on deep optical imaging and the SZ effect in CMB maps, respectively---we constructed a new cross-matched sample of 186 clusters that is larger and deeper than those in many previous miscentering studies. After modeling the distribution of centering offsets in this fiducial sample, we found a miscentered fraction of $\sim 25\%$, a value generally consistent with previous studies. We examined the image of each miscentered cluster in our fiducial sample and identified one of several reasons to explain the miscentering. While some of the miscentered clusters are undergoing mergers, a number of them are miscentered due to non-astrophysical effects: a large fraction of the miscentered clusters can be ascribed to limitations of the HSC data and/or the CAMIRA algorithm, including bright star masks, observational artifacts and deblending issues. We also found a few clusters that are miscentered due to a false match between HSC and ACT clusters or due to an ACT signal that does not correspond to a real cluster. 

In order to further verify our classification of well-centered and miscentered clusters, we computed their weak lensing signals. Compared to well-centered clusters, the miscentered clusters' lensing signal is significantly suppressed, especially at small radii. However, when we computed the miscentered lensing signal with their SZ centers instead of their BCG centers, there was a much clearer signal at small radii. Both of these trends verify our intuition that the BCG centers of miscentered clusters are significantly offset from their true gravitational centers.

We created a cleaned sample of 164 clusters that omits all clusters with clear, non-astrophysical causes of miscentering. The threshold beyond which we define a cluster to be miscentered grows from \ACTrevision{330} kpc to \ACTrevision{370} kpc, and the inferred miscentered fraction is considerably smaller,  $\sim 10\%$. All but one of the clusters for which we were unable to identify a cause for miscentering (\S~\ref{subsubsec:no_cause}) fall below the new threshold, affirming the ability of our cleaned offset model to accurately separate clusters into well-centered and miscentered populations.

Additional work is needed to further investigate how both astrophysical processes and systematic effects in cluster-finding algorithms may contribute to miscentering. Bright, radio-emitting galaxies are not exclusively BCGs \citep{lin_radio_2007}, so it is plausible that radio sources can cause SZ centers to be less accurate. Indeed, such compact sources are known to contaminate ACT cluster data \citep{dicker_observations_2021}, but whether compact sources can induce significant miscentering of the SZ signal needs further investigation. Meanwhile, from a numerical perspective, one could use cluster hydrodynamical simulations \cite[e.g.,][]{gupta_sze_2017, yan_analysis_2020, seppi_offset_2023} that compute both a cluster's optical and gas signals. Such an analysis would allow us to study miscentering without observational effects: for example, one could investigate the fraction of merging clusters that have well-defined centers with a fully-formed central galaxy at the bottom of the cluster potential well. 

We suggest that future iterations of optical cluster-finding algorithms explicitly account for non-astrophysical causes of miscentering. For example, one could assign a ``miscentering probability'' to a cluster based on its proximity to a bright star or observational artifact. One could also develop cluster-finding algorithms that examine optical, X-ray, and/or SZ data simultaneously; such algorithms may reduce the number of false matches between different datasets. Such improvements to optical cluster-finding algorithms may reduce the number of clusters with incorrectly identified optical centers.

By constraining the fraction of clusters that are miscentered due to mergers, one can better understand the history of BCG and cluster formation. Furthermore, one can more precisely correct for miscentering-related biases in cluster lensing analyses for cosmology \citep{mcclintock_dark_2019, zhang_dark_2019}. Another implication of our study is that miscentering is an inevitable systematic in cluster lensing studies---due to ongoing mergers, a nontrivial fraction of clusters will lack any well-defined center. This factor may help motivate cosmological studies that fall outside of conventional cluster cosmology; e.g., studies that derive cosmological parameters from the properties of the most massive galaxies \citep{xhakaj_cluster_2024}.

It is worth noting a few miscentering studies which find different miscentered fractions from ours. For example, \cite{willis_understanding_2021} find a larger miscentered fraction of $\sim 55\%$; this may be due to the fact that their model finds $\sigma_1 = 60$ kpc, which is \ACTrevision{considerably smaller} than our adopted $\sigma_1 = 150$ kpc. Meanwhile, \cite{kelly_dark_2023} find a smaller miscentered fraction of $\sim 10-20\%$. Furthermore, \cite{seppi_offset_2023} uses a different classification system altogether, categorizing clusters as dynamically \textit{relaxed} or \textit{disturbed} instead of well-centered or miscentered, respectively; they find a disturbed fraction of $69\%$. Therefore, it is worthwhile to perform additional miscentering studies in order to place stronger constraints on $f_\text{cen}$. \ACTrevision{Although previous studies have used a two-component model, future studies could explore more complex models; in reality, the division of well-centered and miscentered populations may not be cleanly bimodal.}

With future data releases from HSC and ACT, even larger cross-matched samples can be built to study miscentering. Outside of these two surveys, one might also use upcoming data from the Vera C. Rubin Observatory \citep{ivezic_lsst_2019}, which will be deeper and cover a much larger footprint on the sky; and eROSITA \citep{liu_erosita_2022}, which is mapping the entire sky in X-rays.  

\section*{Acknowledgements}


We would like to thank \ACTrevision{Eduardo Rozo, Tesla Jeltema, Neelima Sehgal, Ed Wollack, and Bruce Partridge} for helpful discussions and suggestions. RD acknowledges support from the NSF Graduate Research Fellowship Program under Grant No.\ DGE-2039656. \ACTrevision{MH acknowledges support from the National Research Foundation of South Africa. CS acknowledges support from the Agencia Nacional de Investigaci\'on y Desarrollo (ANID) through Basal project FB210003.}

The Hyper Suprime-Cam (HSC) collaboration includes the astronomical communities of Japan and Taiwan, and Princeton University. The HSC instrumentation and software were developed by the National Astronomical Observatory of Japan (NAOJ), the Kavli Institute for the Physics and Mathematics of the Universe (Kavli IPMU), the University of Tokyo, the High Energy Accelerator Research Organization (KEK), the Academia Sinica Institute for Astronomy and Astrophysics in Taiwan (ASIAA), and Princeton University. Funding was contributed by the FIRST program from the Japanese Cabinet Office, the Ministry of Education, Culture, Sports, Science and Technology (MEXT), the Japan Society for the Promotion of Science (JSPS), Japan Science and Technology Agency (JST), the Toray Science Foundation, NAOJ, Kavli IPMU, KEK, ASIAA, and Princeton University.

This paper makes use of software developed for Vera C. Rubin Observatory. We thank the Rubin Observatory for making their code available as free software at http://pipelines.lsst.io/.

This paper is based on data collected at the Subaru Telescope and retrieved from the HSC data archive system, which is operated by the Subaru Telescope and Astronomy Data Center (ADC) at NAOJ. Data analysis was in part carried out with the cooperation of Center for Computational Astrophysics (CfCA), NAOJ. We are honored and grateful for the opportunity of observing the Universe from Maunakea, which has the cultural, historical and natural significance in Hawaii.

The Pan-STARRS1 Surveys (PS1) and the PS1 public science archive have been made possible through contributions by the Institute for Astronomy, the University of Hawaii, the Pan-STARRS Project Office, the Max Planck Society and its participating institutes, the Max Planck Institute for Astronomy, Heidelberg, and the Max Planck Institute for Extraterrestrial Physics, Garching, The Johns Hopkins University, Durham University, the University of Edinburgh, the Queen’s University Belfast, the Harvard-Smithsonian Center for Astrophysics, the Las Cumbres Observatory Global Telescope Network Incorporated, the National Central University of Taiwan, the Space Telescope Science Institute, the National Aeronautics and Space Administration under grant No. NNX08AR22G issued through the Planetary Science Division of the NASA Science Mission Directorate, the National Science Foundation grant No. AST-1238877, the University of Maryland, Eotvos Lorand University (ELTE), the Los Alamos National Laboratory, and the Gordon and Betty Moore Foundation.

Support for ACT was through the U.S.~National Science Foundation through awards AST-0408698, AST-0965625, and AST-1440226 for the ACT project, as well as awards PHY-0355328, PHY-0855887 and PHY-1214379. Funding was also provided by Princeton University, the University of Pennsylvania, and a Canada Foundation for Innovation (CFI) award to UBC. ACT operated in the Parque Astron\'omico Atacama in northern Chile under the auspices of the Agencia Nacional de Investigaci\'on y Desarrollo (ANID). The development of multichroic detectors and lenses was supported by NASA grants NNX13AE56G and NNX14AB58G. Detector research at NIST was supported by the NIST Innovations in Measurement Science program. Computing for ACT was performed using the Princeton Research Computing resources at Princeton University, the National Energy Research Scientific Computing Center (NERSC), and the Niagara supercomputer at the SciNet HPC Consortium. SciNet is funded by the CFI under the auspices of Compute Canada, the Government of Ontario, the Ontario Research Fund–Research Excellence, and the University of Toronto. We thank the Republic of Chile for hosting ACT in the northern Atacama, and the local indigenous Licanantay communities whom we follow in observing and learning from the night sky.

\section*{Data Availability}

The HSC optical catalog used in this paper can be found at \url{https://github.com/oguri/cluster_catalogs/tree/main/hsc_s19a_camira}, while the ACT SZ catalog can be found at \url{https://lambda.gsfc.nasa.gov/product/act/actpol_dr5_szcluster_catalog_get.html}. We will provide our fiducial cross-matched catalog upon request.

\bibliographystyle{mnras}
\bibliography{references} 





\appendix

\section{List of Clusters in Fiducial Cross-Match}\label{sec:list_of_cm_clusters}

We provide a full list of clusters in our fiducial cross-match in Table~\ref{tab:fiducial_cm}, including their HSC and ACT positions, redshifts, richnesses, SZ signal-to-noise ratios, measured offsets, whether they are miscentered and/or merging, and the cause of the miscentering.

\scriptsize

\clearpage
\onecolumn

\begin{longtable}{lllllllllllll}
\caption{The full list of clusters used in our fiducial cross-match. All coordinates are listed in degrees. The values for redshift ($z$) and richness ($\lambda$) are taken from the HSC catalog, SNR refers to the ACT SNR, $r$ is the centering offset, ``Miscent.?'' indicates if the cluster is miscentered (see Section~\ref{subsec:well_vs_mis}), ``Merging?'' indicates if the cluster is merging (see Section~\ref{subsec:mergers}), and ``Cause of miscent.'' indicates the cause of miscentering (see Section~\ref{sec:causes_of_miscent})
}\label{tab:fiducial_cm} \\

\hline \multicolumn{1}{|c}{\textbf{HSC name}} & \multicolumn{1}{c}{\textbf{HSC RA ($^\circ$)}} & \multicolumn{1}{c}{\textbf{HSC dec ($^\circ$)}} & \multicolumn{1}{c}{\textbf{ACT name}}& \multicolumn{1}{c}{\textbf{ACT RA ($^\circ$)}}& \multicolumn{1}{c}{\textbf{ACT dec ($^\circ$)}}& \multicolumn{1}{c}{\textbf{$z$}}& \multicolumn{1}{c}{\textbf{$\lambda$}}& \multicolumn{1}{c}{\textbf{SNR}}& \multicolumn{1}{c}{\textbf{$r$ (Mpc)}}& \multicolumn{1}{c}{\textbf{Miscent.?}}& \multicolumn{1}{c}{\textbf{Merging?}}& \multicolumn{1}{c|}{\textbf{Cause of miscent.}} \\ \hline 
\endfirsthead

\multicolumn{13}{c}%
{{\bfseries \tablename\ \thetable{} -- continued from previous page}} \\
\hline \multicolumn{1}{|c}{\textbf{HSC name}} & \multicolumn{1}{c}{\textbf{HSC RA ($^\circ$)}} & \multicolumn{1}{c}{\textbf{HSC dec ($^\circ$)}} & \multicolumn{1}{c}{\textbf{ACT name}}& \multicolumn{1}{c}{\textbf{ACT RA ($^\circ$)}}& \multicolumn{1}{c}{\textbf{ACT dec ($^\circ$)}}& \multicolumn{1}{c}{\textbf{$z$}}& \multicolumn{1}{c}{\textbf{$\lambda$}}& \multicolumn{1}{c}{\textbf{SNR}}& \multicolumn{1}{c}{\textbf{$r$ (Mpc)}}& \multicolumn{1}{c}{\textbf{Miscent.?}}& \multicolumn{1}{c}{\textbf{Merging?}}& \multicolumn{1}{c|}{\textbf{Cause of miscent.}}  \\ \hline 
\endhead

\hline \multicolumn{13}{|r|}{{Continued on next page}} \\ \hline
\endfoot

\hline \hline
\endlastfoot

HSCJ000217+013157 & 0.570211 & 1.532472 & ACT-CL J0002.3+0131 & 0.583333 & 1.524909 & 0.81 & 34.607 & 4.615 & 0.411 & True & False & Multiple possible \\
HSCJ000350+020359 & 0.956952 & 2.066506 & ACT-CL J0003.8+0203 & 0.954179 & 2.06261 & 0.1142 & 31.927 & 5.491 & 0.0357 & False & False & N/A \\
HSCJ000500+021143 & 1.249243 & 2.195339 & ACT-CL J0005.0+0212 & 1.251648 & 2.201674 & 0.827 & 25.975 & 5.04 & 0.1852 & False & False & N/A \\
HSCJ000547+022256 & 1.44459 & 2.382173 & ACT-CL J0005.7+0222 & 1.441633 & 2.370831 & 0.8167 & 24.325 & 4.845 & 0.319 & False & False & N/A \\
HSCJ000634+001806 & 1.641751 & 0.301708 & ACT-CL J0006.5+0017 & 1.643564 & 0.285975 & 0.698 & 30.785 & 5.507 & 0.4069 & True & False & Misidentified CG \\
HSCJ000654-004120 & 1.725384 & -0.688763 & ACT-CL J0006.9-0041 & 1.730549 & -0.686201 & 0.5552 & 59.479 & 9.08 & 0.1337 & False & True & N/A \\
HSCJ000810+020112 & 2.043308 & 2.0201 & ACT-CL J0008.1+0201 & 2.045439 & 2.019734 & 0.354 & 70.37 & 15.931 & 0.0387 & False & True & N/A \\
HSCJ015654-042426 & 29.226423 & -4.407102 & ACT-CL J0156.8-0424 & 29.222169 & -4.405736 & 0.138 & 30.982 & 4.744 & 0.0391 & False & False & N/A \\
HSCJ015826-014639 & 29.607963 & -1.777593 & ACT-CL J0158.4-0145 & 29.603343 & -1.763265 & 0.1625 & 54.471 & 5.527 & 0.1514 & False & False & N/A \\
HSCJ015913-032420 & 29.802687 & -3.405454 & ACT-CL J0159.2-0324 & 29.812338 & -3.408333 & 0.5407 & 22.081 & 4.422 & 0.2301 & False & False & N/A \\
HSCJ015921-030955 & 29.839286 & -3.165221 & ACT-CL J0159.3-0310 & 29.829256 & -3.170783 & 0.622 & 20.469 & 4.719 & 0.2803 & False & False & N/A \\
HSCJ015931-035023 & 29.877133 & -3.839823 & ACT-CL J0159.4-0350 & 29.874309 & -3.839241 & 0.7799 & 67.619 & 5.754 & 0.0771 & False & False & N/A \\
HSCJ020147-021154 & 30.445168 & -2.198314 & ACT-CL J0201.6-0211 & 30.423806 & -2.198919 & 0.204 & 72.431 & 9.219 & 0.2576 & False & True & N/A \\
HSCJ020300-004236 & 30.748478 & -0.709983 & ACT-CL J0203.0-0042 & 30.751043 & -0.707206 & 0.4246 & 31.269 & 5.34 & 0.0758 & False & False & N/A \\
HSCJ020317-012401 & 30.819173 & -1.40022 & ACT-CL J0203.2-0123 & 30.815637 & -1.396675 & 0.7011 & 28.334 & 6.107 & 0.1289 & False & True & N/A \\
HSCJ020447-011713 & 31.196516 & -1.287044 & ACT-CL J0204.7-0116 & 31.190626 & -1.28232 & 0.312 & 29.098 & 4.717 & 0.1244 & False & False & N/A \\
HSCJ020456-030321 & 31.232113 & -3.055755 & ACT-CL J0204.8-0303 & 31.209438 & -3.06144 & 0.5479 & 25.113 & 10.422 & 0.5378 & True & False & Star \\
HSCJ020517-043920 & 31.319592 & -4.655573 & ACT-CL J0205.2-0439 & 31.317053 & -4.653123 & 0.96 & 39.663 & 12.089 & 0.1005 & False & True & N/A \\
HSCJ020613-011500 & 31.554741 & -1.250012 & ACT-CL J0206.2-0114 & 31.556563 & -1.242799 & 0.6834 & 60.863 & 14.43 & 0.1895 & False & True & N/A \\
HSCJ020623-011832 & 31.595229 & -1.308775 & ACT-CL J0206.4-0118 & 31.608295 & -1.308362 & 0.1955 & 39.762 & 4.823 & 0.1525 & False & False & N/A \\
HSCJ020658-011951 & 31.743357 & -1.330933 & ACT-CL J0206.9-0119 & 31.737571 & -1.333333 & 0.5411 & 24.372 & 4.463 & 0.1434 & False & True & N/A \\
HSCJ020816-023724 & 32.067082 & -2.62341 & ACT-CL J0208.2-0237 & 32.064946 & -2.621668 & 0.5124 & 35.454 & 5.581 & 0.0613 & False & False & N/A \\
HSCJ020820-025608 & 32.081581 & -2.935489 & ACT-CL J0208.3-0255 & 32.085929 & -2.93323 & 0.946 & 26.834 & 5.658 & 0.1392 & False & False & N/A \\
HSCJ021002-024411 & 32.507873 & -2.736408 & ACT-CL J0210.0-0243 & 32.504257 & -2.720847 & 0.658 & 23.021 & 4.609 & 0.4006 & True & False & Deblending \\
HSCJ021056-061154 & 32.735023 & -6.198431 & ACT-CL J0210.9-0611 & 32.737387 & -6.193518 & 0.4315 & 43.368 & 5.797 & 0.1102 & False & True & N/A \\
HSCJ021102-045318 & 32.757776 & -4.88833 & ACT-CL J0211.1-0453 & 32.8 & -4.895703 & 0.144 & 15.437 & 4.655 & 0.3885 & True & False & False match \\
HSCJ021115-034324 & 32.811752 & -3.723309 & ACT-CL J0211.2-0343 & 32.811616 & -3.723237 & 0.7536 & 62.216 & 8.936 & 0.0041 & False & False & N/A \\
HSCJ021321-060437 & 33.337991 & -6.076813 & ACT-CL J0213.3-0605 & 33.337377 & -6.099905 & 0.702 & 48.262 & 6.518 & 0.5949 & True & False & Multiple possible \\
HSCJ021426-062740 & 33.607429 & -6.461161 & ACT-CL J0214.4-0628 & 33.613948 & -6.467543 & 0.2308 & 50.971 & 5.219 & 0.1206 & False & False & N/A \\
HSCJ021441-043402 & 33.671241 & -4.567356 & ACT-CL J0214.7-0432 & 33.67881 & -4.548587 & 0.1485 & 68.543 & 4.554 & 0.1888 & False & True & N/A \\
HSCJ021528-044041 & 33.867653 & -4.678115 & ACT-CL J0215.4-0440 & 33.866667 & -4.670873 & 0.356 & 41.609 & 4.435 & 0.1314 & False & False & N/A \\
HSCJ021835-011434 & 34.644288 & -1.242899 & ACT-CL J0218.5-0114 & 34.64254 & -1.242802 & 0.868 & 40.461 & 5.742 & 0.0486 & False & False & N/A \\
HSCJ022056-033348 & 35.231588 & -3.563314 & ACT-CL J0220.9-0332 & 35.236988 & -3.547877 & 1.0139 & 36.94 & 5.137 & 0.4729 & True & False & Multiple possible \\
HSCJ022146-034619 & 35.440625 & -3.771928 & ACT-CL J0221.7-0346 & 35.439659 & -3.772847 & 0.42 & 82.458 & 10.414 & 0.0265 & False & True & N/A \\
HSCJ022156-034054 & 35.481354 & -3.68159 & ACT-CL J0221.9-0340 & 35.481415 & -3.669552 & 1.052 & 32.01 & 6.229 & 0.351 & True & True & Merger \\
HSCJ022528-035502 & 36.364613 & -3.917308 & ACT-CL J0225.4-0355 & 36.366797 & -3.933333 & 0.7846 & 27.499 & 4.564 & 0.4342 & True & False & False signal \\
HSCJ022738-031757 & 36.909256 & -3.299209 & ACT-CL J0227.6-0318 & 36.911256 & -3.30087 & 0.847 & 54.071 & 6.824 & 0.0715 & False & False & N/A \\
HSCJ022936-033623 & 37.401047 & -3.606418 & ACT-CL J0229.6-0336 & 37.410493 & -3.611185 & 0.316 & 37.784 & 6.913 & 0.1755 & False & True & N/A \\
HSCJ023141-045257 & 37.921574 & -4.882613 & ACT-CL J0231.7-0453 & 37.935949 & -4.886 & 0.187 & 117.669 & 9.247 & 0.1658 & False & True & N/A \\
HSCJ023217-061914 & 38.070717 & -6.320418 & ACT-CL J0232.2-0617 & 38.052728 & -6.294417 & 1.1521 & 22.545 & 4.289 & 0.9361 & True & True & Merger \\
HSCJ023336-053022 & 38.398282 & -5.506037 & ACT-CL J0233.6-0530 & 38.405235 & -5.508124 & 0.4296 & 49.864 & 8.612 & 0.1459 & False & True & N/A \\
HSCJ023516-032348 & 38.816422 & -3.396777 & ACT-CL J0235.2-0323 & 38.808404 & -3.387505 & 0.578 & 42.807 & 4.682 & 0.2897 & False & True & N/A \\
HSCJ023643-022953 & 39.180914 & -2.497939 & ACT-CL J0236.7-0228 & 39.179107 & -2.470875 & 0.604 & 18.859 & 4.763 & 0.6548 & True & False & Artifact \\
HSCJ023655-060708 & 39.228944 & -6.11882 & ACT-CL J0236.8-0607 & 39.224926 & -6.125074 & 0.688 & 24.131 & 5.467 & 0.1896 & False & True & N/A \\
HSCJ083932-014128 & 129.884537 & -1.691122 & ACT-CL J0839.5-0140 & 129.876467 & -1.676617 & 0.2605 & 47.413 & 6.536 & 0.2408 & False & False & N/A \\
HSCJ084225+003236 & 130.602173 & 0.54322 & ACT-CL J0842.3+0033 & 130.595658 & 0.554184 & 1.028 & 36.414 & 4.78 & 0.37 & True & True & Merger \\
HSCJ084441+021656 & 131.171072 & 2.282303 & ACT-CL J0844.6+0216 & 131.170545 & 2.276926 & 0.6404 & 69.556 & 5.256 & 0.1339 & False & False & N/A \\
HSCJ084502+012631 & 131.259589 & 1.442025 & ACT-CL J0845.0+0127 & 131.25981 & 1.46531 & 0.4165 & 29.912 & 5.858 & 0.4615 & True & False & Multiple possible \\
HSCJ084526+032732 & 131.35941 & 3.458953 & ACT-CL J0845.4+0327 & 131.36915 & 3.462576 & 0.322 & 69.067 & 12.237 & 0.1746 & False & False & N/A \\
HSCJ084824+041206 & 132.098806 & 4.201732 & ACT-CL J0848.4+0412 & 132.105052 & 4.204553 & 0.8533 & 58.139 & 6.005 & 0.1888 & False & False & N/A \\
HSCJ084831-012137 & 132.127353 & -1.360332 & ACT-CL J0848.5-0122 & 132.13888 & -1.367742 & 0.616 & 16.443 & 4.903 & 0.3337 & True & True & Merger \\
HSCJ084939-005121 & 132.413654 & -0.855933 & ACT-CL J0849.6-0051 & 132.416562 & -0.854322 & 0.616 & 50.99 & 5.684 & 0.081 & False & True & N/A \\
HSCJ085132+045227 & 132.881637 & 4.874083 & ACT-CL J0851.5+0452 & 132.888864 & 4.880626 & 0.4 & 33.919 & 4.36 & 0.1882 & False & False & N/A \\
HSCJ085221-010047 & 133.086607 & -1.012999 & ACT-CL J0852.2-0101 & 133.070604 & -1.020816 & 0.438 & 39.217 & 7.313 & 0.3635 & True & True & Star \\
HSCJ085450-012157 & 133.706648 & -1.365851 & ACT-CL J0854.7-0122 & 133.695815 & -1.37217 & 0.3428 & 27.248 & 5.195 & 0.22 & False & False & N/A \\
HSCJ085530-010656 & 133.875864 & -1.115507 & ACT-CL J0855.4-0106 & 133.869549 & -1.115603 & 0.7491 & 39.263 & 5.664 & 0.1668 & False & True & N/A \\
HSCJ085622+014656 & 134.092204 & 1.782199 & ACT-CL J0856.3+0146 & 134.086063 & 1.783076 & 0.7636 & 34.7 & 8.662 & 0.1649 & False & False & N/A \\
HSCJ085629+014431 & 134.122187 & 1.741871 & ACT-CL J0856.5+0143 & 134.128276 & 1.7257 & 0.7067 & 49.379 & 6.504 & 0.4462 & True & True & Star \\
HSCJ085754+031035 & 134.474797 & 3.176485 & ACT-CL J0857.8+0310 & 134.46401 & 3.178078 & 0.192 & 61.529 & 11.373 & 0.1252 & False & True & N/A \\
HSCJ085903-012004 & 134.760998 & -1.334328 & ACT-CL J0859.0-0119 & 134.75414 & -1.333333 & 1.336 & 26.834 & 4.638 & 0.2095 & False & False & N/A \\
HSCJ085932+030841 & 134.884675 & 3.144733 & ACT-CL J0859.4+0308 & 134.875 & 3.14588 & 0.186 & 26.059 & 4.14 & 0.1091 & False & True & N/A \\
HSCJ090119+030156 & 135.330628 & 3.03219 & ACT-CL J0901.5+0301 & 135.385139 & 3.019336 & 0.194 & 42.298 & 4.844 & 0.6487 & True & True & Star \\
HSCJ090141-013648 & 135.420606 & -1.613274 & ACT-CL J0901.5-0139 & 135.375769 & -1.66332 & 0.3031 & 25.609 & 5.671 & 1.085 & True & True & Artifact \\
HSCJ090330-013622 & 135.875628 & -1.606151 & ACT-CL J0903.4-0136 & 135.871684 & -1.603024 & 0.4448 & 43.979 & 5.458 & 0.1036 & False & False & N/A \\
HSCJ090419+020641 & 136.079294 & 2.111412 & ACT-CL J0904.3+0206 & 136.079874 & 2.106649 & 0.8076 & 28.802 & 4.698 & 0.1301 & False & False & N/A \\
HSCJ090754+005732 & 136.976449 & 0.958979 & ACT-CL J0907.8+0057 & 136.971935 & 0.953086 & 0.686 & 31.757 & 5.153 & 0.1894 & False & True & N/A \\
HSCJ090756+025422 & 136.981336 & 2.906203 & ACT-CL J0907.9+0254 & 136.983333 & 2.912502 & 0.8073 & 46.372 & 4.106 & 0.1791 & False & False & N/A \\
HSCJ090932-005016 & 137.385315 & -0.837881 & ACT-CL J0909.5-0050 & 137.383377 & -0.835399 & 1.09 & 38.623 & 4.606 & 0.0925 & False & True & N/A \\
HSCJ091449+001018 & 138.704735 & 0.171563 & ACT-CL J0914.7+0011 & 138.695729 & 0.191667 & 0.5414 & 18.816 & 4.253 & 0.5045 & True & False & Star \\
HSCJ091555-013244 & 138.981014 & -1.545651 & ACT-CL J0915.9-0132 & 138.989099 & -1.545581 & 0.5 & 33.561 & 5.778 & 0.1776 & False & True & N/A \\
HSCJ091606-002338 & 139.023871 & -0.393884 & ACT-CL J0916.1-0024 & 139.04253 & -0.401908 & 0.288 & 60.628 & 10.714 & 0.3165 & False & True & N/A \\
HSCJ091850+021203 & 139.709235 & 2.200894 & ACT-CL J0918.7+0211 & 139.686703 & 2.190027 & 0.266 & 54.374 & 4.464 & 0.3682 & True & True & Merger \\
HSCJ092024+013444 & 140.098056 & 1.578875 & ACT-CL J0920.4+0134 & 140.100017 & 1.577392 & 0.68 & 51.074 & 5.61 & 0.0625 & False & False & N/A \\
HSCJ092121+031713 & 140.337964 & 3.286995 & ACT-CL J0921.3+0317 & 140.335619 & 3.288125 & 0.3513 & 105.961 & 10.708 & 0.0463 & False & True & N/A \\
HSCJ092211+034641 & 140.545649 & 3.77819 & ACT-CL J0922.1+0346 & 140.533808 & 3.773285 & 0.25 & 65.172 & 5.249 & 0.1801 & False & True & N/A \\
HSCJ093512+004932 & 143.801208 & 0.825611 & ACT-CL J0935.2+0048 & 143.80869 & 0.800777 & 0.3521 & 78.408 & 13.913 & 0.463 & True & True & Merger \\
HSCJ093523+023325 & 143.843843 & 2.556878 & ACT-CL J0935.3+0233 & 143.842726 & 2.551368 & 0.5214 & 46.617 & 11.436 & 0.1263 & False & True & N/A \\
HSCJ094345+005916 & 145.936227 & 0.987781 & ACT-CL J0943.7+0059 & 145.947342 & 0.988996 & 0.4343 & 34.384 & 4.681 & 0.2271 & False & True & N/A \\
HSCJ094747-013154 & 146.94549 & -1.531617 & ACT-CL J0947.7-0132 & 146.94504 & -1.534906 & 0.882 & 30.164 & 5.069 & 0.0926 & False & False & N/A \\
HSCJ094759-011941 & 146.994134 & -1.328169 & ACT-CL J0947.9-0120 & 146.992928 & -1.334111 & 1.1041 & 54.343 & 13.207 & 0.1785 & False & True & N/A \\
HSCJ095140-001420 & 147.917065 & -0.238932 & ACT-CL J0951.6-0014 & 147.921976 & -0.239948 & 0.4271 & 45.255 & 9.527 & 0.1009 & False & False & N/A \\
HSCJ095612+015935 & 149.04878 & 1.992959 & ACT-CL J0956.1+0158 & 149.044615 & 1.980276 & 0.946 & 43.346 & 5.947 & 0.3796 & True & True & Merger \\
HSCJ100452-013748 & 151.214658 & -1.629998 & ACT-CL J1004.8-0137 & 151.212664 & -1.629311 & 0.3017 & 25.549 & 5.475 & 0.0339 & False & True & N/A \\
HSCJ100532+023518 & 151.381352 & 2.588443 & ACT-CL J1005.5+0235 & 151.383407 & 2.585845 & 0.372 & 49.045 & 4.741 & 0.0612 & False & False & N/A \\
HSCJ100944+025413 & 152.43191 & 2.903656 & ACT-CL J1009.7+0255 & 152.428211 & 2.926406 & 0.382 & 39.706 & 6.556 & 0.4333 & True & False & Multiple possible \\
HSCJ101229+023958 & 153.122707 & 2.666196 & ACT-CL J1012.5+0239 & 153.129263 & 2.662599 & 0.24 & 81.221 & 8.224 & 0.102 & False & True & N/A \\
HSCJ101402+023754 & 153.509016 & 2.631569 & ACT-CL J1014.0+0237 & 153.504148 & 2.625 & 0.5773 & 27.166 & 4.126 & 0.1932 & False & True & N/A \\
HSCJ101407+003827 & 153.530422 & 0.64089 & ACT-CL J1014.1+0038 & 153.529466 & 0.642712 & 1.2433 & 44.502 & 10.402 & 0.0617 & False & False & N/A \\
HSCJ101443+031037 & 153.679031 & 3.176877 & ACT-CL J1014.6+0310 & 153.666752 & 3.180976 & 0.7381 & 34.751 & 4.643 & 0.3395 & True & True & Merger \\
HSCJ101846-013103 & 154.690919 & -1.517421 & ACT-CL J1018.7-0130 & 154.692935 & -1.516609 & 0.3953 & 29.722 & 6.614 & 0.0417 & False & True & N/A \\
HSCJ102006-001306 & 155.025417 & -0.218415 & ACT-CL J1020.1-0013 & 155.027719 & -0.219368 & 0.3939 & 57.07 & 4.577 & 0.0477 & False & False & N/A \\
HSCJ102750+000328 & 156.957222 & 0.057738 & ACT-CL J1027.8+0004 & 156.958334 & 0.06895 & 0.714 & 27.956 & 5.27 & 0.2921 & False & False & N/A \\
HSCJ102956+001650 & 157.483609 & 0.280606 & ACT-CL J1029.9+0016 & 157.478109 & 0.277933 & 1.347 & 27.972 & 6.001 & 0.1851 & False & False & N/A \\
HSCJ103112+003130 & 157.800549 & 0.525109 & ACT-CL J1031.2+0031 & 157.808297 & 0.529233 & 0.8811 & 17.087 & 4.88 & 0.2447 & False & False & N/A \\
HSCJ103525-000141 & 158.855791 & -0.028051 & ACT-CL J1035.4-0001 & 158.85925 & -0.032014 & 0.4824 & 68.922 & 5.859 & 0.1134 & False & True & N/A \\
HSCJ104713-011612 & 161.806053 & -1.270103 & ACT-CL J1047.2-0116 & 161.802374 & -1.270221 & 0.7439 & 43.676 & 4.73 & 0.0969 & False & True & N/A \\
HSCJ104723-010430 & 161.847253 & -1.074965 & ACT-CL J1047.4-0103 & 161.850941 & -1.065627 & 0.478 & 26.515 & 5.327 & 0.2153 & False & False & N/A \\
HSCJ105040+001707 & 162.666271 & 0.285331 & ACT-CL J1050.6+0017 & 162.667243 & 0.284428 & 0.59 & 28.162 & 7.019 & 0.0317 & False & False & N/A \\
HSCJ110728+012138 & 166.867214 & 1.360674 & ACT-CL J1107.4+0121 & 166.860813 & 1.35588 & 0.5872 & 47.358 & 5.122 & 0.1905 & False & True & N/A \\
HSCJ111017-003023 & 167.572548 & -0.506422 & ACT-CL J1110.2-0030 & 167.573192 & -0.504919 & 0.9954 & 31.428 & 5.167 & 0.0471 & False & True & N/A \\
HSCJ111111+004508 & 167.796717 & 0.752179 & ACT-CL J1111.1+0044 & 167.798448 & 0.748426 & 0.1971 & 49.759 & 5.667 & 0.0485 & False & True & N/A \\
HSCJ111654-003648 & 169.223668 & -0.613324 & ACT-CL J1116.9-0036 & 169.228957 & -0.612512 & 0.8338 & 45.79 & 4.687 & 0.1466 & False & False & N/A \\
HSCJ111846+003851 & 169.691587 & 0.647626 & ACT-CL J1118.7+0039 & 169.691618 & 0.659881 & 0.588 & 40.423 & 5.58 & 0.2922 & False & False & N/A \\
HSCJ112549+001017 & 171.452369 & 0.171272 & ACT-CL J1125.7+0010 & 171.431499 & 0.178289 & 0.748 & 42.912 & 4.855 & 0.5811 & True & True & Merger \\
HSCJ112733+000342 & 171.885477 & 0.061779 & ACT-CL J1127.5+0003 & 171.891548 & 0.066667 & 1.3641 & 17.599 & 4.159 & 0.2361 & False & False & N/A \\
HSCJ112818-005900 & 172.07378 & -0.983255 & ACT-CL J1128.3-0058 & 172.075036 & -0.981827 & 0.468 & 72.298 & 6.399 & 0.0403 & False & True & N/A \\
HSCJ113139-010718 & 172.913695 & -1.121702 & ACT-CL J1131.6-0107 & 172.91175 & -1.123247 & 0.321 & 68.464 & 5.912 & 0.0417 & False & False & N/A \\
HSCJ113408+011536 & 173.532139 & 1.259885 & ACT-CL J1134.0+0116 & 173.512546 & 1.27917 & 0.736 & 31.233 & 5.013 & 0.7211 & True & True & Merger \\
HSCJ113654+000643 & 174.226799 & 0.112083 & ACT-CL J1136.9+0006 & 174.231486 & 0.100596 & 0.59 & 79.803 & 14.27 & 0.2963 & False & True & N/A \\
HSCJ113833-001137 & 174.639459 & -0.193605 & ACT-CL J1138.5-0011 & 174.638836 & -0.197338 & 0.8821 & 47.315 & 4.417 & 0.1055 & False & False & N/A \\
HSCJ113919+015412 & 174.827091 & 1.903244 & ACT-CL J1139.3+0154 & 174.831111 & 1.904592 & 1.0527 & 36.16 & 8.141 & 0.1236 & False & False & N/A \\
HSCJ114409+044133 & 176.039345 & 4.692382 & ACT-CL J1144.0+0440 & 176.025 & 4.679106 & 0.4093 & 16.596 & 4.129 & 0.3827 & True & False & False match \\
HSCJ114416+022722 & 176.068573 & 2.456173 & ACT-CL J1144.2+0227 & 176.054194 & 2.462539 & 0.6512 & 45.159 & 4.652 & 0.3923 & True & True & Merger \\
HSCJ114507-010032 & 176.280823 & -1.008907 & ACT-CL J1145.1-0100 & 176.287483 & -1.004184 & 0.6999 & 30.793 & 4.186 & 0.21 & False & False & N/A \\
HSCJ115216+003055 & 178.066216 & 0.51537 & ACT-CL J1152.2+0031 & 178.056037 & 0.523485 & 0.4597 & 80.716 & 8.073 & 0.2732 & False & True & N/A \\
HSCJ115235+035641 & 178.145684 & 3.944594 & ACT-CL J1152.5+0356 & 178.144056 & 3.94412 & 0.708 & 46.783 & 9.799 & 0.0437 & False & True & N/A \\
HSCJ115328+025849 & 178.368111 & 2.980402 & ACT-CL J1153.4+0259 & 178.370846 & 2.987527 & 0.72 & 19.31 & 5.107 & 0.1985 & False & True & N/A \\
HSCJ115417+022126 & 178.571393 & 2.357235 & ACT-CL J1154.2+0221 & 178.56668 & 2.360253 & 0.7266 & 43.832 & 10.862 & 0.146 & False & True & N/A \\
HSCJ115823+022806 & 179.593864 & 2.468256 & ACT-CL J1158.3+0228 & 179.597372 & 2.470706 & 0.49 & 25.459 & 8.518 & 0.093 & False & False & N/A \\
HSCJ120054-012826 & 180.224356 & -1.473969 & ACT-CL J1200.9-0125 & 180.231264 & -1.433293 & 0.246 & 22.295 & 4.445 & 0.5739 & True & True & Merger \\
HSCJ120350+012523 & 180.958005 & 1.423187 & ACT-CL J1203.9+0126 & 180.9872 & 1.438063 & 0.4181 & 42.186 & 7.641 & 0.6507 & True & True & Star \\
HSCJ120533+021425 & 181.386577 & 2.240404 & ACT-CL J1205.5+0214 & 181.383333 & 2.245762 & 0.7747 & 41.578 & 4.347 & 0.1674 & False & True & N/A \\
HSCJ120827+025640 & 182.113731 & 2.944318 & ACT-CL J1208.3+0254 & 182.087571 & 2.916667 & 0.7909 & 20.104 & 4.37 & 1.0242 & True & False & Multiple possible \\
HSCJ121018+022344 & 182.573797 & 2.395467 & ACT-CL J1210.2+0223 & 182.574889 & 2.391678 & 0.3942 & 28.761 & 6.012 & 0.0756 & False & False & N/A \\
HSCJ121015+032116 & 182.564078 & 3.354489 & ACT-CL J1210.2+0321 & 182.55591 & 3.364153 & 1.0324 & 42.004 & 6.952 & 0.3672 & True & False & None apparent \\
HSCJ121322+033226 & 183.340625 & 3.540642 & ACT-CL J1213.4+0331 & 183.356343 & 3.529743 & 0.3312 & 17.78 & 5.07 & 0.3276 & False & True & N/A \\
HSCJ121340+035120 & 183.418209 & 3.855565 & ACT-CL J1213.6+0351 & 183.416808 & 3.854246 & 0.8659 & 37.534 & 5.163 & 0.0533 & False & True & N/A \\
HSCJ121832+023328 & 184.633529 & 2.557662 & ACT-CL J1218.5+0233 & 184.62618 & 2.554919 & 0.768 & 59.801 & 9.37 & 0.2088 & False & False & N/A \\
HSCJ122226-012712 & 185.610265 & -1.45325 & ACT-CL J1222.4-0127 & 185.608345 & -1.452562 & 0.294 & 44.692 & 5.206 & 0.0322 & False & True & N/A \\
HSCJ122422+021211 & 186.091575 & 2.20301 & ACT-CL J1224.3+0211 & 186.079273 & 2.199995 & 0.4444 & 63.077 & 4.899 & 0.2606 & False & False & N/A \\
HSCJ123835-000359 & 189.646202 & -0.066417 & ACT-CL J1238.5-0004 & 189.643246 & -0.066732 & 0.588 & 36.675 & 5.26 & 0.0709 & False & False & N/A \\
HSCJ124100+001101 & 190.250186 & 0.183654 & ACT-CL J1241.0+0010 & 190.255022 & 0.176655 & 0.7904 & 33.969 & 4.426 & 0.229 & False & False & N/A \\
HSCJ125036+003646 & 192.64933 & 0.612847 & ACT-CL J1250.6+0036 & 192.651492 & 0.601483 & 0.6851 & 56.503 & 4.821 & 0.295 & False & False & N/A \\
HSCJ125825+004417 & 194.603949 & 0.738087 & ACT-CL J1258.4+0043 & 194.620819 & 0.733333 & 0.4 & 31.104 & 4.666 & 0.339 & True & False & None apparent \\
HSCJ131131-011933 & 197.880946 & -1.325761 & ACT-CL J1311.5-0119 & 197.878621 & -1.33201 & 0.1836 & 111.773 & 24.339 & 0.074 & False & False & N/A \\
HSCJ131504+012113 & 198.766517 & 1.353512 & ACT-CL J1315.0+0121 & 198.766667 & 1.354123 & 0.738 & 32.825 & 4.179 & 0.0165 & False & True & N/A \\
HSCJ134614-014307 & 206.557026 & -1.718662 & ACT-CL J1346.2-0142 & 206.558458 & -1.71314 & 1.1899 & 55.088 & 6.26 & 0.1701 & False & True & N/A \\
HSCJ134745+012410 & 206.936073 & 1.402849 & ACT-CL J1347.7+0124 & 206.936216 & 1.412002 & 0.556 & 21.375 & 5.861 & 0.2124 & False & False & N/A \\
HSCJ135710-011106 & 209.290027 & -1.185015 & ACT-CL J1357.1-0111 & 209.283527 & -1.191765 & 0.396 & 51.516 & 6.963 & 0.1801 & False & False & N/A \\
HSCJ135746+002431 & 209.441211 & 0.408675 & ACT-CL J1357.6+0026 & 209.408333 & 0.437492 & 0.67 & 17.306 & 4.376 & 1.1047 & True & False & False signal \\
HSCJ140707-001505 & 211.779312 & -0.251406 & ACT-CL J1407.1-0015 & 211.781773 & -0.256624 & 0.57 & 22.5 & 9.649 & 0.1355 & False & True & N/A \\
HSCJ140803+012048 & 212.011534 & 1.34675 & ACT-CL J1407.9+0120 & 211.983214 & 1.339616 & 0.461 & 23.666 & 4.36 & 0.6137 & True & False & Star \\
HSCJ141558+011809 & 213.99351 & 1.302471 & ACT-CL J1415.9+0118 & 213.994344 & 1.305542 & 0.8246 & 20.084 & 4.569 & 0.0869 & False & False & N/A \\
HSCJ142017+005645 & 215.069493 & 0.945796 & ACT-CL J1420.2+0057 & 215.073566 & 0.95954 & 0.51 & 38.496 & 5.932 & 0.3183 & False & False & N/A \\
HSCJ142025-004309 & 215.104968 & -0.71928 & ACT-CL J1420.4-0042 & 215.114975 & -0.715447 & 0.7927 & 28.787 & 6.316 & 0.2887 & False & False & N/A \\
HSCJ142032+002737 & 215.134538 & 0.460402 & ACT-CL J1420.5+0027 & 215.138519 & 0.451265 & 0.6306 & 89.632 & 7.597 & 0.2453 & False & True & N/A \\
HSCJ142103+002322 & 215.260753 & 0.389554 & ACT-CL J1421.0+0022 & 215.262208 & 0.370932 & 0.654 & 49.918 & 5.76 & 0.4672 & True & True & Merger \\
HSCJ142438+010915 & 216.158045 & 1.154275 & ACT-CL J1424.6+0108 & 216.15 & 1.137449 & 0.622 & 36.917 & 4.129 & 0.4563 & True & True & False signal \\
HSCJ142456-014152 & 216.234116 & -1.697835 & ACT-CL J1424.9-0141 & 216.238952 & -1.686031 & 0.948 & 37.315 & 4.502 & 0.3629 & True & False & None apparent \\
HSCJ142624-012657 & 216.601055 & -1.449167 & ACT-CL J1426.3-0126 & 216.594728 & -1.447166 & 0.4528 & 76.091 & 8.103 & 0.138 & False & False & N/A \\
HSCJ143607-005405 & 219.027581 & -0.901277 & ACT-CL J1436.1-0054 & 219.037497 & -0.908114 & 0.69 & 36.61 & 5.065 & 0.308 & False & True & N/A \\
HSCJ144031-013732 & 220.130763 & -1.625652 & ACT-CL J1440.5-0137 & 220.126279 & -1.624222 & 0.3088 & 89.282 & 8.639 & 0.077 & False & True & N/A \\
HSCJ144133-005358 & 220.38613 & -0.899519 & ACT-CL J1441.5-0053 & 220.390202 & -0.890215 & 0.5372 & 49.524 & 4.58 & 0.2317 & False & False & N/A \\
HSCJ144309+010105 & 220.788791 & 1.018041 & ACT-CL J1443.1+0102 & 220.79119 & 1.041791 & 0.5287 & 31.387 & 6.359 & 0.5401 & True & False & Star \\
HSCJ144955+005036 & 222.480008 & 0.843226 & ACT-CL J1450.0+0049 & 222.523337 & 0.824991 & 0.382 & 28.364 & 5.259 & 0.8838 & True & True & Star \\
HSCJ145219+001027 & 223.078667 & 0.174028 & ACT-CL J1452.3+0009 & 223.07932 & 0.162355 & 0.597 & 40.924 & 6.874 & 0.2807 & False & True & N/A \\
HSCJ221016+045118 & 332.566339 & 4.854976 & ACT-CL J2210.1+0451 & 332.537481 & 4.854123 & 0.578 & 36.453 & 4.288 & 0.6803 & True & False & Star \\
HSCJ221359+015127 & 333.497898 & 1.857536 & ACT-CL J2213.9+0151 & 333.491937 & 1.854227 & 0.702 & 57.824 & 5.401 & 0.1755 & False & True & N/A \\
HSCJ221552+043503 & 333.965633 & 4.584258 & ACT-CL J2215.8+0435 & 333.966651 & 4.589261 & 0.6355 & 85.625 & 5.835 & 0.1261 & False & True & N/A \\
HSCJ222435+050716 & 336.143852 & 5.121022 & ACT-CL J2224.5+0507 & 336.141515 & 5.12084 & 0.622 & 23.496 & 5.061 & 0.0571 & False & True & N/A \\
HSCJ222524+044832 & 336.347955 & 4.808806 & ACT-CL J2225.4+0447 & 336.353053 & 4.79926 & 0.5 & 40.588 & 4.858 & 0.2376 & False & False & N/A \\
HSCJ222808+023325 & 337.033003 & 2.557033 & ACT-CL J2228.0+0233 & 337.024901 & 2.562467 & 0.7835 & 18.588 & 4.968 & 0.2616 & False & False & N/A \\
HSCJ222936+031840 & 337.398028 & 3.311074 & ACT-CL J2229.5+0318 & 337.397435 & 3.309464 & 0.3006 & 26.702 & 4.67 & 0.0275 & False & True & N/A \\
HSCJ223504+033326 & 338.766167 & 3.557252 & ACT-CL J2235.0+0333 & 338.762807 & 3.550872 & 0.7381 & 60.172 & 8.556 & 0.1893 & False & True & N/A \\
HSCJ225238+043142 & 343.15845 & 4.528324 & ACT-CL J2252.6+0432 & 343.150459 & 4.537319 & 0.696 & 103.433 & 8.278 & 0.3084 & False & True & N/A \\
HSCJ230235+000234 & 345.6461 & 0.042839 & ACT-CL J2302.6+0002 & 345.651516 & 0.041885 & 0.534 & 75.622 & 6.556 & 0.1251 & False & True & N/A \\
HSCJ230335+000855 & 345.895069 & 0.148492 & ACT-CL J2303.5+0008 & 345.887408 & 0.141518 & 0.5159 & 61.962 & 4.656 & 0.2314 & False & True & N/A \\
HSCJ230740+013056 & 346.916265 & 1.515519 & ACT-CL J2307.6+0130 & 346.91726 & 1.515662 & 0.4053 & 73.259 & 10.455 & 0.0196 & False & False & N/A \\
HSCJ231255-001521 & 348.229546 & -0.255778 & ACT-CL J2312.9-0013 & 348.22911 & -0.233333 & 0.588 & 35.433 & 4.207 & 0.5352 & True & False & None apparent \\
HSCJ231459+023150 & 348.746425 & 2.530622 & ACT-CL J2314.8+0231 & 348.724994 & 2.520886 & 0.5812 & 26.991 & 4.417 & 0.5577 & True & True & Star \\
HSCJ231833+003403 & 349.638652 & 0.567605 & ACT-CL J2318.5+0033 & 349.629097 & 0.562501 & 0.7386 & 32.976 & 4.598 & 0.2845 & False & True & N/A \\
HSCJ231950+003029 & 349.957945 & 0.508105 & ACT-CL J2319.7+0030 & 349.948159 & 0.501628 & 0.8901 & 26.936 & 7.892 & 0.3281 & False & True & N/A \\
HSCJ231953+003813 & 349.972636 & 0.637046 & ACT-CL J2319.8+0038 & 349.968109 & 0.644234 & 0.872 & 30.935 & 5.237 & 0.236 & False & True & N/A \\
HSCJ232618+003002 & 351.574507 & 0.50051 & ACT-CL J2326.2+0030 & 351.570636 & 0.505178 & 1.1799 & 40.416 & 5.652 & 0.1806 & False & False & N/A \\
HSCJ232915+005653 & 352.312526 & 0.94813 & ACT-CL J2329.2+0056 & 352.321084 & 0.937561 & 0.742 & 25.08 & 4.663 & 0.3578 & True & False & Merger \\
HSCJ233239-001350 & 353.163728 & -0.2305 & ACT-CL J2332.6-0014 & 353.171805 & -0.247035 & 1.092 & 31.045 & 5.678 & 0.5407 & True & True & Misidentified CG \\
HSCJ233253+011050 & 353.218979 & 1.180664 & ACT-CL J2332.8+0110 & 353.215821 & 1.167041 & 0.48 & 52.484 & 6.939 & 0.3006 & False & True & N/A \\
HSCJ233331-005630 & 353.378333 & -0.941535 & ACT-CL J2333.4-0056 & 353.371849 & -0.943868 & 0.5168 & 25.108 & 4.973 & 0.1541 & False & False & N/A \\
HSCJ233740+001617 & 354.415545 & 0.27138 & ACT-CL J2337.6+0016 & 354.412551 & 0.272715 & 0.2878 & 88.112 & 23.567 & 0.0511 & False & False & N/A \\
HSCJ234311+003231 & 355.794545 & 0.541862 & ACT-CL J2343.1+0032 & 355.787502 & 0.541497 & 0.8504 & 40.002 & 6.666 & 0.1945 & False & False & N/A \\
HSCJ234341+001845 & 355.920075 & 0.312438 & ACT-CL J2343.7+0018 & 355.9282 & 0.306792 & 0.2554 & 104.584 & 7.651 & 0.1415 & False & True & N/A \\
HSCJ234736+011602 & 356.899439 & 1.267264 & ACT-CL J2347.6+0116 & 356.902542 & 1.268125 & 1.266 & 36.758 & 7.864 & 0.0968 & False & True & N/A \\
HSCJ235127+004633 & 357.864107 & 0.775861 & ACT-CL J2351.4+0047 & 357.872388 & 0.784359 & 0.436 & 41.036 & 5.146 & 0.2416 & False & False & N/A \\
HSCJ235145+000916 & 357.936213 & 0.154526 & ACT-CL J2351.7+0009 & 357.938857 & 0.153106 & 0.8588 & 38.777 & 8.349 & 0.083 & False & False & N/A \\
HSCJ235628+001028 & 359.116487 & 0.174372 & ACT-CL J2356.3+0009 & 359.097946 & 0.158315 & 1.0545 & 26.853 & 4.42 & 0.7156 & True & True & Deblending \\
HSCJ235934+020824 & 359.889826 & 2.139925 & ACT-CL J2359.5+0208 & 359.887357 & 2.138787 & 0.4303 & 64.955 & 9.27 & 0.0549 & False & False & N/A \\

\end{longtable}

\clearpage
\twocolumn

\normalsize

\section{Properties of Merging vs. Non-Merging Clusters}\label{sec:props_merg_vs_non}

\begin{figure*}
\centering
\includegraphics[width=0.95\textwidth]{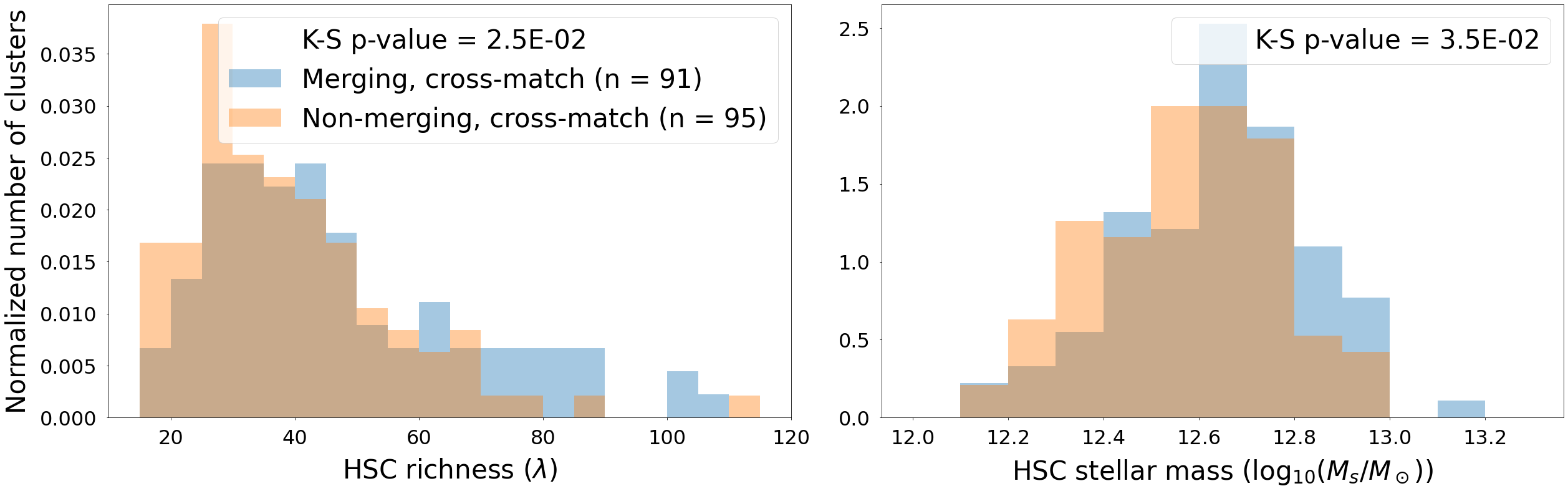}
     \caption[Properties of merging clusters vs. non-merging clusters within the fiducial cross-match]{The normalized distributions of richness and stellar mass for both merging and non-merging clusters within the fiducial cross-match, along with the corresponding K-S test p-values. In this plot, merging and non-merging clusters are categorized based on the merger catalog of \cite{okabe_halo_2019}. The p-values for both $\lambda$ and $M_s$ are statistically significant, indicating that merging clusters tend to have higher richnesses and stellar masses than non-merging clusters. 
     }
     \label{fig:props_merg_non_cm}
\end{figure*}

Fig.~\ref{fig:props_merg_non_cm} compares the distributions of $\lambda$ and $M_s$ between merging and non-merging clusters within the fiducial cross-match. Similar to the well-centered vs. miscentered comparison in Fig.~\ref{fig:props_well_mis}, the merging and non-merging distributions for both properties are inconsistent with being drawn from the same underlying distribution. Fig.~\ref{fig:props_merg_non_cm} also suggests that merging clusters have higher $\lambda$ and $M_s$ than non-merging ones. The same effect is seen in the richness distribution for {\em all} HSC clusters, rather than just the sample in the fiducial cross-match. This is likely a selection effect since clusters with fewer galaxies are less likely to have multiple discernible peaks in their galaxy distribution and thus are less likely to be identified as mergers by \cite{okabe_halo_2019}. 

\bsp	
\label{lastpage}
\end{document}